\documentclass[12pt]{iopart}
\usepackage{graphicx}
\usepackage[font=small,labelfont=bf,
   justification=justified,
   format=plain]{caption}
\usepackage{subcaption}
\usepackage{tikz}
\usetikzlibrary{calc}
\usepackage{hyperref}
\usepackage{wrapfig}
\usepackage{svg}
\usepackage{hhline}
\usepackage{diagbox}
\usepackage{adjustbox}

\usepackage[symbol]{footmisc}

\begin{document}

\title{Fast-response hot-wire flow sensors for wind and gust estimation on UAVs}

\author{Nathaniel Simon$^1$, Alexander Piqué$^1$, David Snyder, \\ Kyle Ikuma, Anirudha Majumdar, and Marcus Hultmark.}

\address{Department of Mechanical and Aerospace Engineering, Princeton University, Princeton, NJ.}
\ead{nsimon@princeton.edu}
\vspace{0.1 in}
\address{$^1$Equal Contribution.}

\begin{abstract}

Due to limitations in available sensor technology, unmanned aerial vehicles (UAVs) lack an active sensing capability to measure turbulence, gusts, or other unsteady aerodynamic phenomena. Conventional \textit{in situ} anemometry techniques fail to deliver in the harsh and dynamic multirotor environment due to form factor, resolution, or robustness requirements. To address this capability gap, a novel, fast-response sensor system to measure a wind vector in two dimensions is introduced and evaluated. This system, known as `MAST' (for MEMS Anemometry Sensing Tower), leverages advances in microelectromechanical (MEMS) hot-wire devices to produce a solid-state, lightweight, and robust flow sensor suitable for real-time wind estimation onboard a UAV. The MAST uses five pentagonally-arranged microscale hot-wires to determine the wind vector's direction and magnitude. The MAST's performance was evaluated in a wind tunnel at speeds up to 5~m/s and orientations of $0^\circ-360^\circ$. A neural network sensor model was trained from the wind tunnel data to estimate the wind vector from sensor signals. The average error of the sensor is 0.14~m/s for speed and 1.6$^\circ$ for direction. Furthermore, 95\% of measurements are within 0.36~m/s for speed and 5.0$^\circ$ for direction. With a bandwidth of 570~Hz determined from square-wave testing, the MAST stands to greatly enhance UAV wind estimation capabilities and enable capturing relevant high-frequency phenomena in flow conditions.

\end{abstract}

\vspace{2pc}

\submitto{\MST}

%
%
\newpage
\section{Introduction}

In real-world flight conditions, turbulence and other unsteady fluid mechanics phenomena necessitate measuring a fluid velocity vector for real-time control or sensing directly. This includes detection of a gust impinging on an unmanned aerial vehicle (UAV), airspeed sensing for fast maneuvering aircraft, or even load predictions for wind turbines. The objective of this study is to improve flow sensing onboard a UAV, which poses a challenging environment for conventional flow sensors. The UAV environment is harsh: subject to rapid accelerations and vibrations. It is also dynamic, meaning that the flow conditions are unsteady and evolve rapidly (e.g. wind gusts and rotor wakes). Furthermore, there are design and integration challenges, such as meeting power, volume, and weight constraints. As UAV technology matures, there is a push to utilize UAVs for increasingly critical applications, ranging from package delivery \cite{frachtenberg2019practical} to infrastructure inspection \cite{shakhatreh2019unmanned} to search-and-rescue operations \cite{schedl2021autonomous}. A major obstacle to such widespread deployment is the variability of weather conditions, such as high winds and unpredictable gusts, that can cancel missions or worse, cause mid-flight accidents. Strong winds and gusts can perturb the drone from its desired trajectory and increase the probability of collision \cite{banerjee2022probability}. To increase drone viability in such environments by managing the uncertainty and unpredictability of wind behavior, drones should be capable of sensing wind in real-time. If the UAV can accurately estimate the relative wind vector, it is able to better anticipate and react to changes in its flight trajectory. Conventional anemometry techniques, however, are not up to the task.

\subsection{Conventional Anemometry Techniques} \label{subsubsec:anemometry-techniques}

One of the most common anemometers is the Pitot-static tube. A Pitot-static tube measures flow speed using the pressure difference between two locations: one at the stagnation point and the other downstream to measure the static pressure \cite{anderson2007fund}. However, Pitot-static tubes have several limitations rendering them generally unsuitable in unsteady flow conditions. These limitations include time lags due to the pressure line's resistance to a traveling pressure signal \cite{weissberg1953response} and susceptibility to fluctuations in pressure lines leading to false interpretations \cite{care2020investigation}. Due to such limitations, settling times on the order of tens of seconds are needed \cite{bailey2013obtaining}, but only mean value measurements are realizable. Furthermore, conventional one-dimensional Pitot-static tubes are not designed to resolve the flow velocity vector's orientation, and are insensitive to misalignment of up to $15-20^\circ$ \cite{beck2010aerodynamics}. In short, while the Pitot-static tube is well-suited as an airspeed indicator on fixed-wing platforms or for measuring velocity in steady flows, its form factor (long tube) and temporal limitations make the Pitot-static tube an ill-suited anemometer in unsteady environments where flow direction is important.

Another classic flow measurement device is the hot-wire anemometer \cite{bruun1995hot}. When a heated thermistor (i.e. the hot-wire) is exposed to flow, convective cooling reduces the temperature of the wire and thus its resistance. The change in resistance is then related to a velocity. Since the hot-wire operating principle relies on heat transfer, it generally has a faster response than the Pitot-static tube with time constants on the order of milliseconds (microscale wires) or microseconds (nanoscale wires) \cite{bailey2010turbulence}. However, the hot-wire also has its limitations: conventional driving circuits are large, heavy and expensive, which is exacerbated by the necessity of multiple (at least 3) sensing elements to resolve a 3-component velocity field \cite{devenport1992turbulence}, an increase in spatial filtering with additional sensing elements \cite{fan2015nanoscale}, and the intrinsic fragility of the unshielded microscale wire \cite{huang2021investigation}. In short, the fragile nature of hot-wires, as well as their form factor, render them an unsuitable choice for harsh environments such as a flying vehicle.

In summary, conventional \textit{in situ} flow measurement techniques, such as Pitot-static tubes and hot-wires, do not fit the form factor, temporal resolution, or robustness requirements for applications in harsh and dynamic environments where knowledge of the wind direction is critical. Hovering UAVs (e.g., helicopters, multirotors, and hybrid UAVs), regularly operate in, and can themselves generate, harsh and dynamic environments. Among other causes, rotors introduce additional vibration and dynamics (e.g., blade flapping \cite{craig2020geometric}, tip vortices \cite{seddon2011basic}). The hovering task is especially vulnerable to wind gusts, as hovering vehicles have less momentum and disturbances from any direction require immediate correction. For these reasons, hovering UAVs stand to benefit from the integration of novel flow sensors that overcome the inadequacies of conventional anemometers.

\subsection{Flow Detection for UAVs} \label{subsubsec:custom-flow-sensors}

Due to the aforementioned integration challenges, the prevailing approach to wind estimation on UAVs is to avoid flow sensors altogether and instead utilize existing on-board sensors (such as inertial measurement units and GPS) to indirectly estimate the wind speed \cite{tomic2016flying, sikkel2016novel, allison2019estimating}. However, this method is inherently reactive and relies on deviations from the UAV's desired flight path to estimate wind.

In pursuit of instantaneous \textit{in situ} measurement, there have been numerous efforts to design specialized sensors for the UAV task. One such sensing modality relates sensor deflection to flow velocity. For example, a PDMS ribbon \cite{sundin2021soft} or a Hall effect sensor coupled with a resisting plate \cite{zahran2018new}, can be used to measure velocity through deflection. In \cite{sundin2021soft}, FEM simulations predicted a sensor bandwidth of 480~Hz while in \cite{zahran2018new}, there was no mention of the sensor's temporal resolution. In both cases, a single deflection sensor was unable to determine the flow direction. In addition, a study by \cite{tagliabue2020touch} used the deflection of foam fins to estimate the velocity vector, with a maximum root mean square error of 0.38~m/s, but without a discussion of the temporal resolution. Despite the shortcomings of Pitot-static tubes and pressure-based sensors, \cite{yeo2015onboard} and \cite{bruschi2016wind} used differential pressure probes to resolve the wind vector. The errors in resolving flow direction were as high as 15\% of full-scale deflection in \cite{yeo2015onboard} and 12$^\circ$ in \cite{bruschi2016wind}. Neither study explicitly reports the sensor's error in speed estimation nor the sensor's temporal resolution. Fast-response multi-hole pressure probes (MHPP) are an established laboratory anemometry technique capable of measuring turbulence up to 2~kHz. These systems, however, are typically expensive and require bulky data acquisition devices and regular calibrations. An effort to extend MHPP performance to the UAV flow sensing problem yielded the Rapid Prototyped Probe, with a 90$^\circ$ cone of acceptance, 400~Hz bandwidth, and measurement accuracy of 1~m/s and 5$^\circ$ \cite{prudden2017anemometer}. Disadvantages of this approach included the inability to measure outside of the cone of acceptance, and the necessary vehicle modifications to accommodate the large probe (which was the length of the vehicle itself) \cite{prudden2018measuring}.

In summary, flow sensor selection for UAVs is a trade-off: physical robustness versus temporal resolution. Pitot-static tubes are the most robust of conventional anemometry techniques, but have temporal resolutions on the order of seconds, whereas nanoscale hot-wires have microsecond resolution, but lack robustness. Therefore, conventional anemometry techniques are ill-equipped for the next generation of drones. Recent aforementioned efforts have produced novel sensors to estimate wind velocity in UAV applications. These studies produced encouraging results with usable performance. However, there is limited or nonexistent discussion regarding the temporal resolution of their novel sensor-model packages. Lacking sufficient temporal resolution, these sensors filter out high-frequency information such as features introduced by turbulence, fast maneuvers, or other unsteady effects. 

The flow length scales in the lower atmosphere typically range from millimeters to kilometers, with time scales as small as tens of milliseconds near the surface \cite{watkins2010low}. These timescales are similarly applicable in complex terrain, such as above a tree canopy \cite{marshall2002conditional}, in a valley \cite{letson2019characterizing}, or in an urban site \cite{rakib2020measured}. For a vehicle moving against the flow the timescale requirements can be significantly more stringent. For example, to detect a flow feature of two centimeters moving at a relative airspeed of 10 m/s, a sensing bandwidth of 500~Hz is required. Outfitted with such a fast-response sensor, a drone could ultimately be deployed as a sensor itself, for example in wind farms to measure the wakes of wind turbines. In conclusion, flow sensors with a fast temporal resolution and accurate wind direction detection are needed to improve UAV performance in gusty conditions as well as to enable UAV-borne anemometers in field campaigns.

\subsection{Statement of Contributions} \label{subsec:statement-of-contributions}

Taking inspiration from classical hot-wire anemometry (HWA) operating principles, developments in microelectromechanical systems (MEMS) flow sensing devices have enabled a new class of solid-state, robust, and fast-response flow sensors \cite{fan2015nanoscale, fu2016elastic}. This novel sensor, referred to in this paper as the ``MEMS Hotwire'', shows promise for UAV and other harsh and dynamic flow sensing applications. Here, the small natural timescales of MEMS devices are leveraged in a highly robust sensor, with a $-3$~dB open-loop bandwidth of 570~Hz as determined by square-wave testing. This is sufficient to temporally resolve most of the aerodynamic phenomena of interest, and the simplicity of the operating circuit makes it a very low cost and lightweight sensor system.

Hotwire anemometry is anisotropic, meaning that it is sensitive to flow direction. A combination of several MEMS Hotwire sensors can leverage their anisotropy to simultaneously determine the direction and magnitude of an incoming wind or gust. (This same approach is used by crossed hot-wires, though they are typically restricted to a cone of acceptance $\le 180^\circ$  \cite{jorgensen2001measure}.) In order to test multiple configurations of the MEMS Hotwire, a modular mounting platform called MAST (MEMS Anemometry Sensing Tower) was designed. Of the tested combinations, the pentagon MAST (comprised of five pentagonally-arranged MEMS Hotwires) was identified to have the best overall performance due to its high accuracy in determining both the magnitude and direction of the wind. The MAST and its accompanying sensor model (a neural network) are able to estimate a wind vector with an average accuracy 0.14~m/s for speed and 1.6$^\circ$ for direction. The sensor error's empirical 95\%-confidence upper bound is 0.36~m/s and 5.0$^\circ$, respectively. With its high accuracy and bandwidth, this sensor is poised to enhance flow measurement capabilities onboard UAVs.

In this paper, Section \ref{sec:mems-hotwire} describes the theory of operation of the MEMS Hotwire. The MAST is described in Section \ref{sec:MAST}. Section \ref{sec:methodology} introduces the experimental facilities and instrumentation used to validate the sensor's performance. In Section \ref{subsec:speed-angle-calibration}, the MAST's sensitivity to flow speed and angle is discussed. The bandwidth of the sensor is evaluated by measuring its step response to a square wave input in Section \ref{sec:frequency-response}.  Finally, a neural network model to estimate wind velocity from the analog outputs of the MAST is proposed and its sensitivity and accuracy is reported in Section \ref{sec:wind-model}. The resulting sensor system is ready for integration on a UAV platform or other task, as discussed in Section \ref{sec:uav_integration} and demonstrated in Figure \ref{fig:flowdrone_integration}.

\section{Theory of Operation} \label{sec:theory-of-operation}

\subsection{MEMS Hotwire} \label{sec:mems-hotwire}

\begin{figure}
\begin{subfigure}[b]{.5\textwidth}
  \centering
  \adjustbox{Clip=0cm 0.9cm 2.8cm 0cm}{%
  \includegraphics[width=1.5\linewidth]{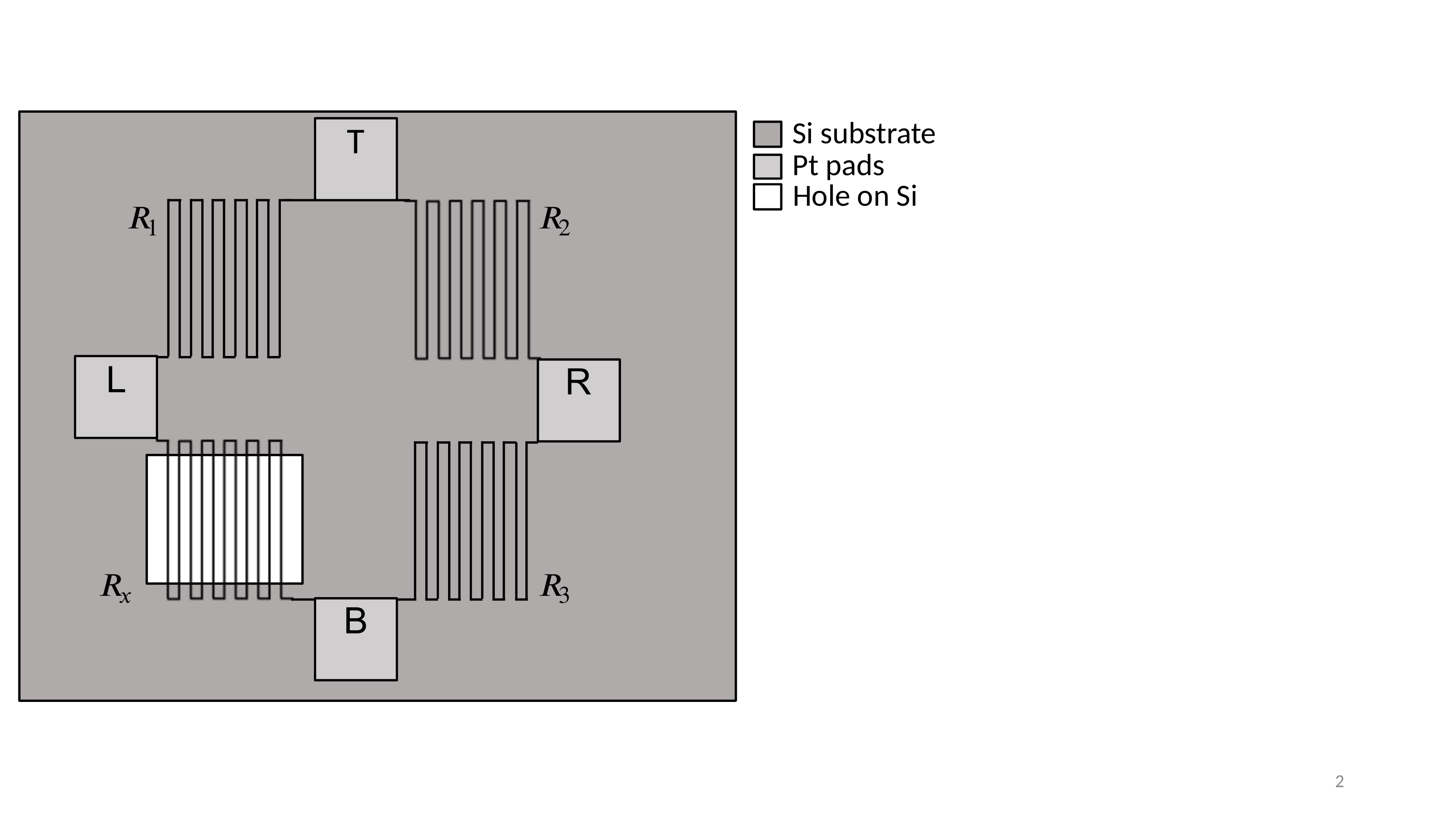}
  }
  \caption{MEMS Hotwire Schematic}
  \label{fig:sfig1}
\end{subfigure}%
\begin{subfigure}[b]{.5\textwidth}
  \centering
    \begin{tikzpicture}
        \node[above right, inner sep=0] (image) at (0,0) {
            \includegraphics[width=.8\linewidth]{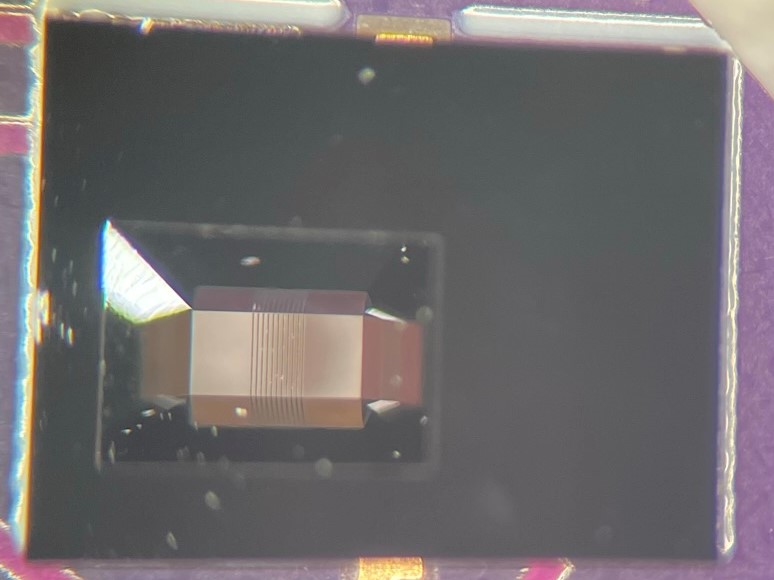}

        };
        \begin{scope}[
        x={($0.1*(image.south east)$)},
        y={($0.1*(image.north west)$)}]

            \draw[latex-, very thick,red] (4.5,4) -- ++ (1.0,0.5)
                node[above right,black,fill=white]{\small flow-through};

        \end{scope}
    \end{tikzpicture}
  \caption{MEMS Hotwire Image}
  \label{fig:sfig2}
\end{subfigure}
\caption{ (a) MEMS Hotwire Schematic as a Wheatstone bridge with leg resistances $R_1, R_2, R_3, R_x$. A window in the silicon substrate allows air to flow through the L-B leg. Convection causes $R_x$ to differ in temperature (and thus resistance) from $R_1,R_2,R_3$, which is measured as a voltage difference across the L and R pads given by Equation \ref{eqn:wheatstone}. (b) A photograph of the MEMS Hotwire under microscope with silicon flow-through visible.}
\label{fig:schematic}
\end{figure}

The MEMS Hotwire is a $4.2 \times 3.3 \times 0.66$mm silicon substrate with four wire arrays arranged as legs of a Wheatstone bridge, as depicted in Figure \ref{fig:schematic}. Each wire array is comprised of a platinum ribbon, with a cross section of 10 x 0.1~$\mu$m, doubled back multiple times such that each array consists of eleven parallel platinum ribbons. The platinum is deposited onto a silicon substrate, which is electrically insulated by a thin layer of silicon nitride.  The resulting wire arrays act as temperature-sensitive resistors in the Wheatstone bridge. A constant voltage is supplied to the top of the bridge, generating heat via Joule heating in all four legs. Like a typical hot-wire, the sensing mode of the MEMS Hotwire depends on a relative difference in convective cooling (due to the flow) between one leg and the others. In this case, the Bottom-Left sensing ribbons (as depicted in Figure \ref{fig:schematic}) are made freestanding using silicon wet etch techniques. This results in a rectangular opening in the silicon through which the air flows, exposing the free standing array to convective cooling from the air. The difference in temperature and thus resistance between the freestanding wire array and the three embedded wire arrays creates a voltage difference across the Left (L) and Right (R) nodes which is measured as the signal of MEMS Hotwire. The MEMS Hotwire dies were provided by Tendo Technologies, Inc.

The equation for the resulting signal can be derived from Kirchhoff's circuit laws:

\begin{equation}
    V_L - V_R = \left( \frac{R_x}{R_x + R_1 }- \frac{R_3}{R_2+R_3} \right)V_T.
    \label{eqn:wheatstone}
\end{equation}
At zero velocity, $ V_L - V_R>0$ since $R_x > R_{1,2,3}$, as the three embedded arrays are being cooled through conduction to the silicon. Assuming that the temperature of the silicon heat sink (and thus the embedded resistors) remains constant and equal to the ambient temperature, as flow velocity increases and $R_x$ cools and drops, the difference $R_x - R_{1,2,3}$ decreases and thus the signal $ V_L - V_R$ decreases. This indicates the existence of an upper limit in measurable velocity (as $R_x - R_{1,2,3} \rightarrow 0$), which depends on the voltage applied to the bridge.

Two design choices contribute to the simplicity of operation and small packaging of the MEMS Hotwire. First, all four legs of the Wheatstone bridge are present on the silicon substrate and thus similarly subjected to the conditions (e.g., ambient temperature) of the environment. Second, the MEMS Hotwire is designed to be operated in constant-voltage (CVA) rather than constant-current (CCA) or constant-temperature (CTA) modes. While CVA operation reduces the dynamic range of the sensor (due to an upper voltage limit set by sensor burnout, much like CCA), the resulting circuit does not need feedback to operate; this results in a very simple and stable operating circuitry.

\subsection{The MAST} \label{sec:MAST}

\begin{figure}
\begin{subfigure}[t]{0.5\textwidth}
  \centering
    \begin{tikzpicture}
        \node[above right, inner sep=0] (image) at (0,0) {
            \includegraphics[trim={2cm 0 1.5cm 4cm},clip,width=0.6\textwidth]{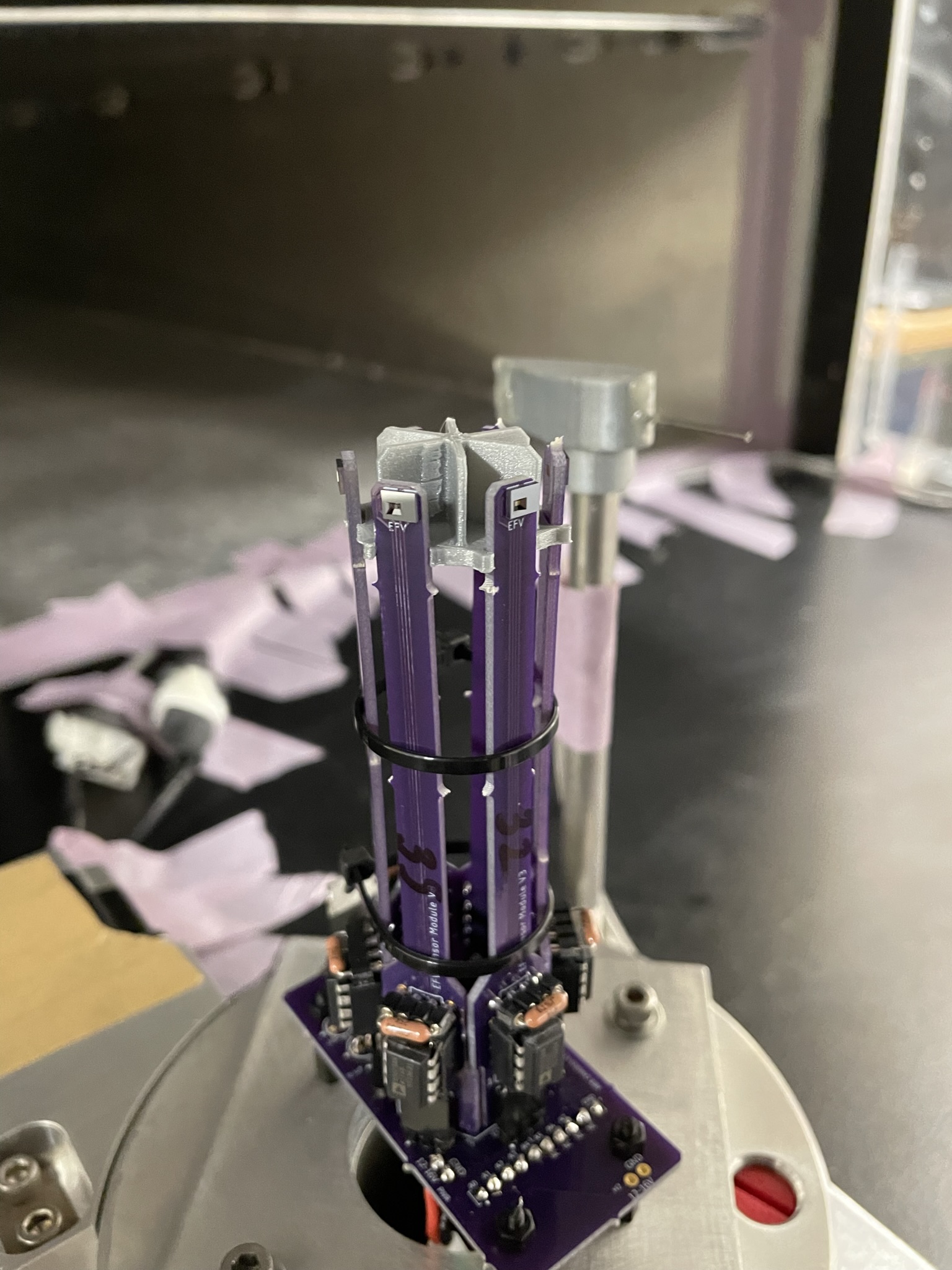}
        };
        \begin{scope}[
        x={($0.1*(image.south east)$)},
        y={($0.1*(image.north west)$)}]
         
            \draw[latex-, very thick,white] (7.5,8) -- ++ (1,-1)
                node[below,black,fill=white]{\small Pitot tube};
            \draw[latex-, very thick,white] (4,8.5) -- ++ (-.5,0.5)
                node[above left,black,fill=white]{\small X-Brace};
            \draw[latex-, very thick,white] (3.25,7.75) -- ++ (-1,0)
                node[left,black,fill=white]{\small MEMS Hotwire};
            \draw[latex-, very thick,white] (3,4.5) -- ++ (-1,-0.5)
                node[below left,black,fill=white]{\small pole-PCB};
            \draw[latex-, very thick,white] (3.5,1) -- ++ (-1,0)
                node[left,black,fill=white]{\small chassis-PCB};
        \end{scope}
    \end{tikzpicture}
    \caption{Hexagon MAST Architecture}
    \label{fig:hexagon_MAST}
\end{subfigure}
\begin{subfigure}[t]{0.5\textwidth}
  \centering
    \begin{tikzpicture}
        \node[above right, inner sep=0] (image) at (0,0) {
                \includegraphics[width=0.6\linewidth]{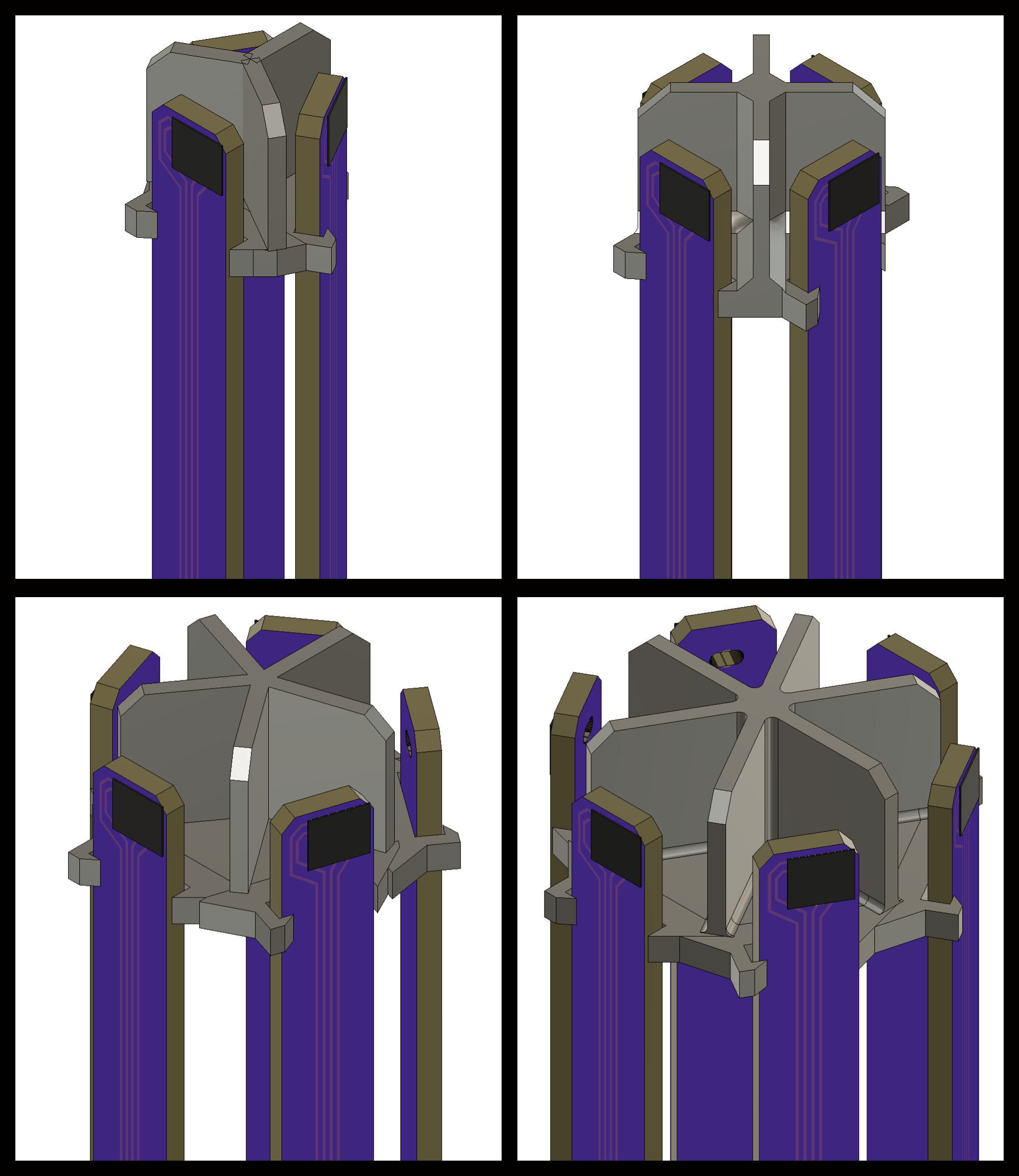}
        };
        \begin{scope}[
        x={($0.1*(image.south east)$)},
        y={($0.1*(image.north west)$)}]
         
            \node[above] at (2.5,10) {Triangle};
            \node[above] at (7.5,10) {Square};
            \node[below] at (2.5,0) {Pentagon};
            \node[below] at (7.5,0) {Hexagon};

        \end{scope}
    \end{tikzpicture}
    \caption{MAST Comparison (CAD)}
    \label{fig:MAST_comparison_CAD}
\end{subfigure}
\caption{ (a) The Hexagon MAST on the rotary stage in the wind tunnel. It is comprised of a single chassis-PCB and six perpendicular pole-PCBs. A Pitot tube is used for ground-truth velocity measurement. The MEMS Hotwire sensors are visible at the top of the pole-PCBs. A 3D printed ``X-Brace'' provides stability and prevents crossflow from opposing sensors. (b) CAD rendering of the four MASTs evaluated in this studyEach MAST had the same height (106~mm) and the hexagon had the greatest width (35~mm). The MAST's width is also its spatial resolution, since the entire MAST is used to calculate the wind estimate. In other words, the MAST cannot resolve velocity gradients within its width.}
\label{fig:MAST}
\end{figure}
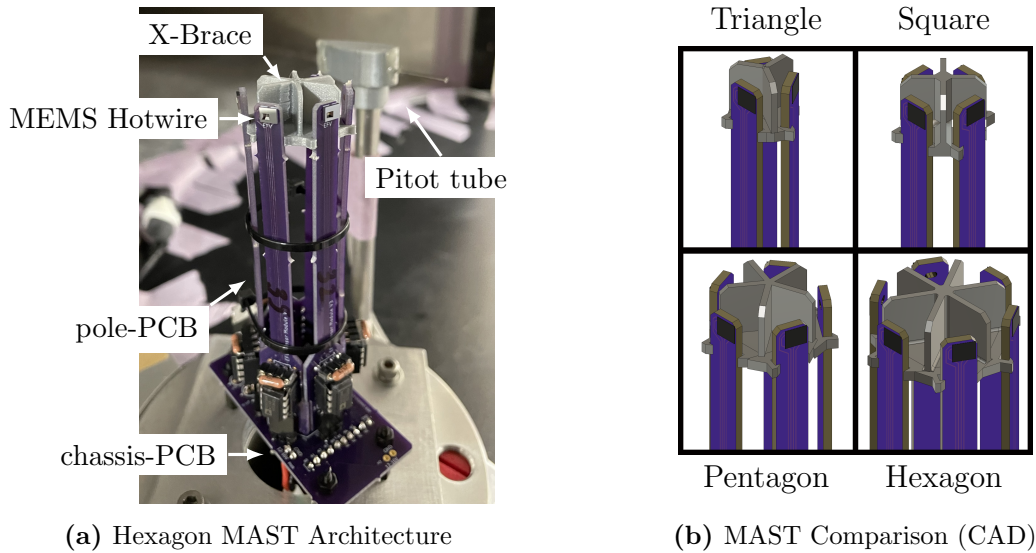

The MAST (MEMS anemometry sensing tower) places the MEMS Hotwire sensors in the desired flow region and configuration to enable meaningful measurement of the magnitude and direction of the wind. There are two main components of the MAST: the \textit{pole-PCB} (printed circuit board), on which a single MEMS Hotwire is placed, and a \textit{chassis-PCB}, on which several pole-PCBs are mounted orthogonally. Figure \ref{fig:hexagon_MAST} depicts six pole-PCBs mounted on a chassis-PCB.

The pole-PCB serves to place and orient an individual MEMS Hotwire sensor in the desired location and orientation. A Tresky Flip Chip Bonder is used to bond the MEMS Hotwire to the pole-PCB with silver epoxy. Once bonded, the fragile sensing element (freestanding platinum wires) is protected between the PCB and the silicon substrate, improving the robustness of the design. An instrument amplifier (AD623) on each pole-PCB amplifies the signal for greater resolution.

The chassis-PCB provides a high-precision constant 10~V source via an AD587 to each of the pole-PCBs. The pole-PCBs are connected orthogonally to the chassis-PCB via right-angle header connectors. The location of header pins on the chassis-PCB determines the number and orientations of pole-PCBs. Since it is intuitive that increasing the number of sensors will monotonically improve performance, a self-imposed design constraint was to determine the fewest number of MEMS Hotwire sensors needed to \textit{sufficiently} resolve the wind vector. The criteria for sufficiency are discussed in Section \ref{sec:wind-model}. Due to the MAST's modular architecture, rapid iteration and testing of the regular polygon MAST designs shown in Figure \ref{fig:MAST_comparison_CAD} was possible. Section \ref{sec:methodology} describes how the MEMS Hotwire and each MAST design was characterized through wind tunnel and frequency response testing.

\section{Experimental Setup} \label{sec:methodology}
Two experimental studies were conducted. First, the MAST's response to wind speed was evaluated in a wind tunnel. To measure how the MAST responded to a change in wind direction, the MAST was placed on a rotary stage to change the relative angle of the MAST to the incoming flow. The wind tunnel facility and corresponding instrumentation are discussed in Section \ref{subsec:wind_tunnel}. Second, the temporal response of an individual MEMS Hotwire was evaluated using a square-wave test; this technique is commonly used to measure the temporal resolution of conventional hot-wires. The square-wave instrumentation is discussed in Section \ref{sec:squareWaveInstrumentation}.

\subsection{Wind Tunnel Facility} \label{subsec:wind_tunnel}

\begin{figure}
\begin{subfigure}[b]{.4\textwidth}
  \centering
    \includegraphics[height=5cm]{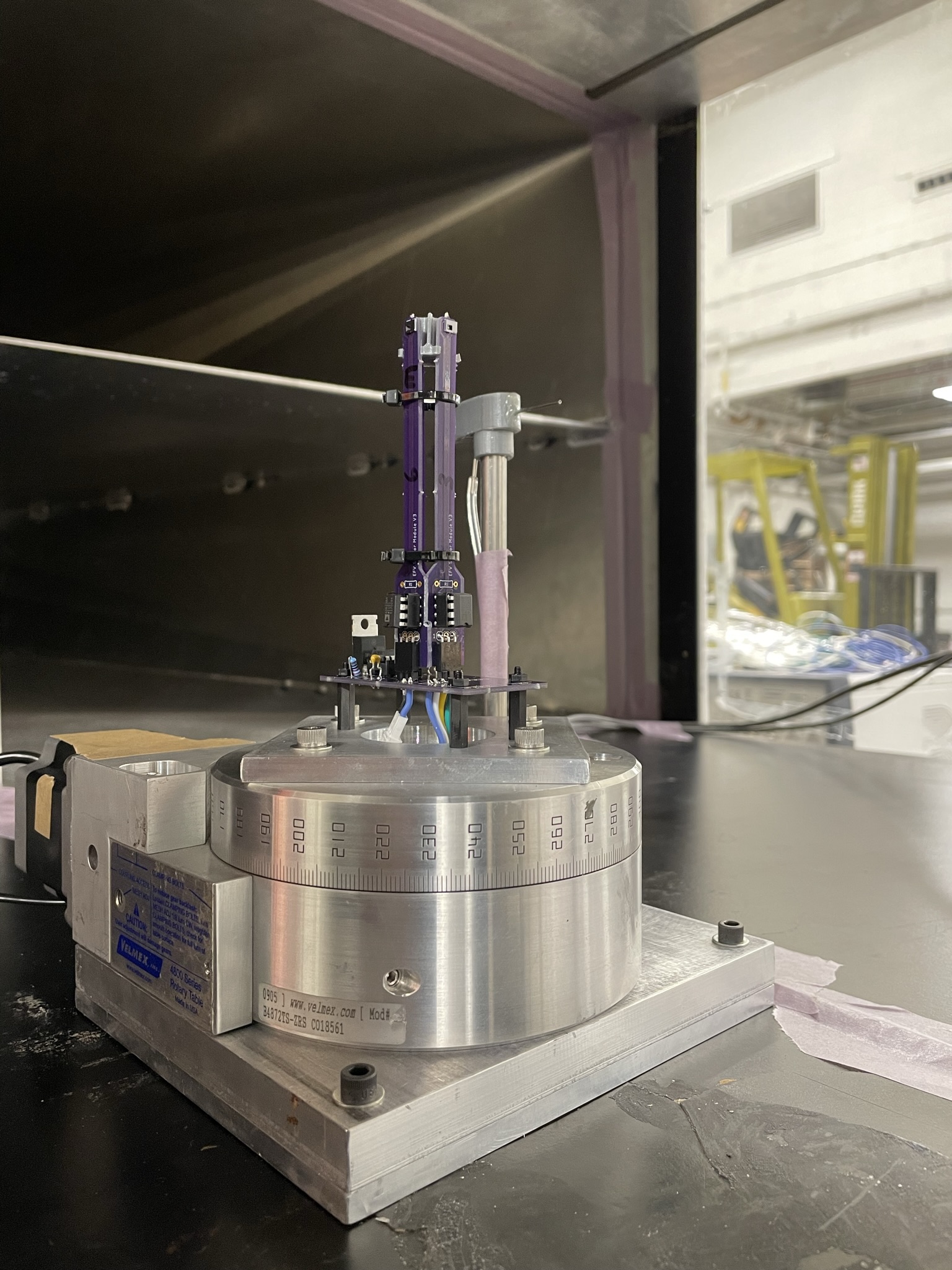}
  \caption{}
  \label{fig:rotary_table}
\end{subfigure}%
\begin{subfigure}[b]{.4\textwidth}
  \centering
    \includegraphics[scale=0.4]{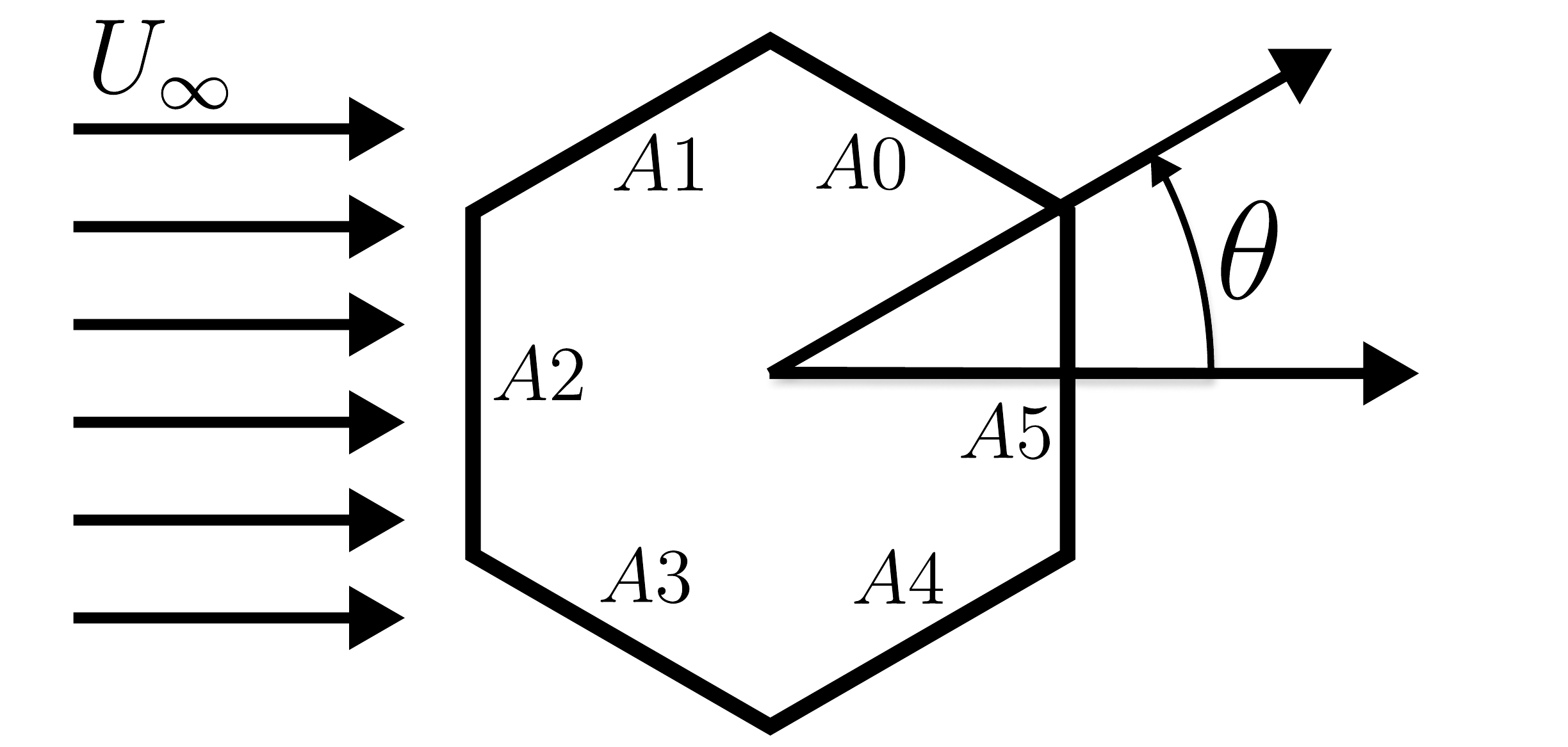} 
  \caption{}
  \label{fig:MASTorientation}
\end{subfigure}
\caption{(a) Rotary table with a square MAST. (b) Labeling of MEMS Hotwire sensors on the hexagon MAST and MAST orientation with respect to the freestream velocity, $U_{\infty}$. Positive $\theta$ represents a counterclockwise rotation from the horizontal axis.}
\end{figure}

The MAST was mounted on a Velmex B4800TS Rotary Table, shown in Figure \ref{fig:rotary_table}. The MAST and rotary table were placed in an open-return wind tunnel measuring $1.2 \times 0.6$~m in cross section. A Pitot tube as depicted in Figure \ref{fig:hexagon_MAST} provided a velocity reference. The velocities measured by the Pitot tube were used to calibrate the MAST voltage measurements to flow velocities. For data acquisition, an NI\textsuperscript{TM} Data Acquisition (DAQ) device (PCI-6341) was used. For all data presented here, measurements taken with no flow are subtracted from the raw voltage data to appropriately zero the non-zero velocity measurements. In addition, the data presented here has been compensated for a gain of 5 from the instrumentation amplifier located on the pole-PCB. In other words, the voltages plotted in Section \ref{sec:results} are pre-amplification. To evaluate the MAST's sensitivity to wind magnitude, eight different freestream velocities, $1.3$~m/s~$\le U_{\infty}\le5.0$~m/s, were tested. At each velocity, the MAST's sensitivity to angle was evaluated by rotating the MAST from $0^\circ\le\theta\le358^\circ$ in 2$^\circ$ increments. The orientation of the MAST with respect to the incoming flow can be found in Figure \ref{fig:MASTorientation}. For each velocity and angle combination, data was sampled at 1 kHz for 30 seconds. A ten second pause between angle changes and a 60 second pause between speed changes were employed to ensure transient effects had dissipated before data collection. This process was repeated for the four different polygon MASTs shown in Figure  \ref{fig:MAST_comparison_CAD}.

\subsection{Temporal Response Instrumentation} \label{sec:squareWaveInstrumentation}

 A common method of measuring a hot-wire's temporal resolution is to determine the sensor's response to a square wave input \cite{freymuth1967feedback,bailey2010turbulence}. Since the MEMS Hotwire sensor has the same operating principle as a hot-wire, the same test was used. In the square wave test, a single pole-PCB was subjected to a 10~V offset $+$ 0.1~V square wave step input at 10~Hz. These parameters were chosen as 10~V is the normal operational voltage of the MEMS Hotwire, and a 0.1~V step input produced a response on the order of a typical velocity change. The step input was applied to the top of the MEMS Hotwire Wheatstone bridge by a Tektronix AFG3000C Arbitrary/Function Generator. The output signal was sampled by the NI\textsuperscript{TM} DAQ at 50kHz. A total sampling duration of 30 seconds yielded 300 periods which were phase-averaged.

\section{Results}\label{sec:results}

\subsection{Angle and speed calibration of the MAST} \label{subsec:speed-angle-calibration}

\begin{figure}[t]
\begin{subfigure}{.45\textwidth}
  \centering
    \includegraphics[width=2.1\linewidth]{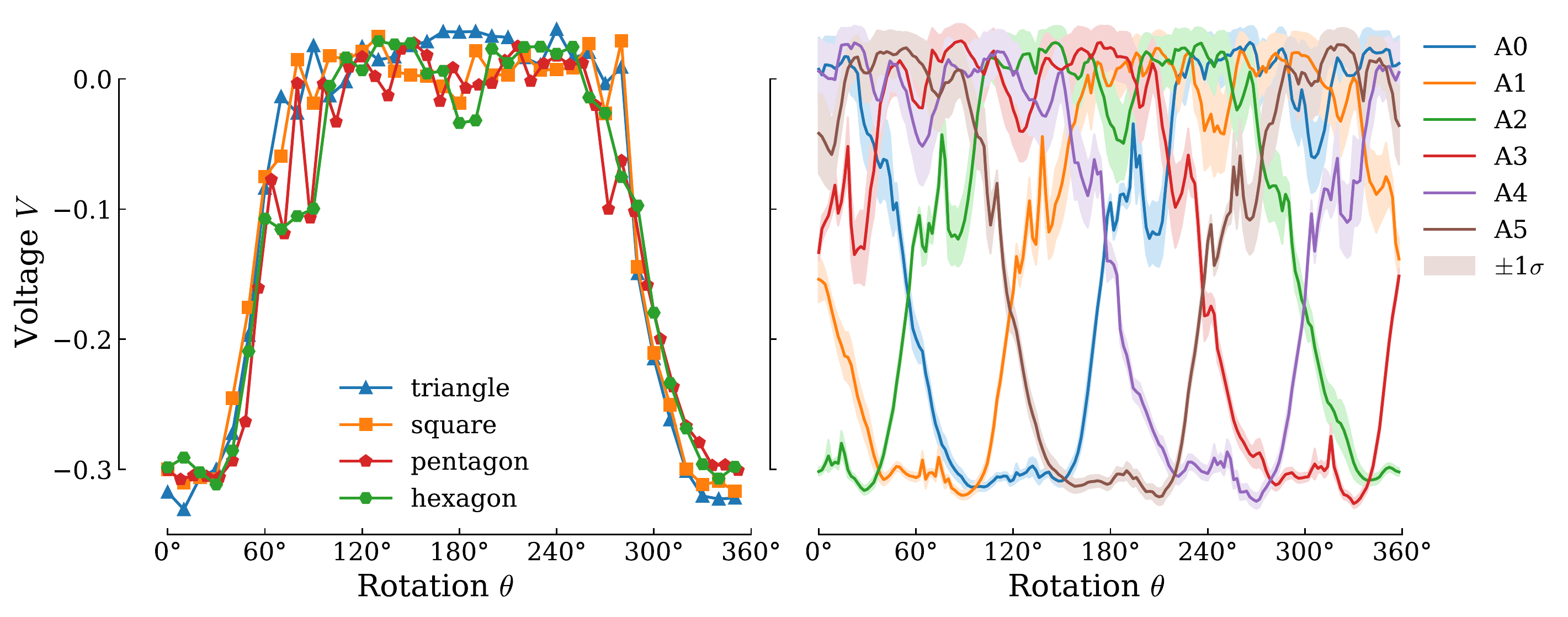}
  \caption{}
  \label{fig:subplot_left}
\end{subfigure}%
\begin{subfigure}[][-158pt][t]{.45\textwidth}
  \centering
  \caption{}
  \label{fig:subplot_right}
\end{subfigure}
\caption{Voltage V vs. MAST rotation $\theta$ for $U_\infty=4.9$ m/s. (a) Angular response of A2, the pole-PCB perpendicular to the flow at $0^\circ$ for each MAST geometry. (b) The six analog outputs of the hexagon MAST. A2 on the Right matches the hexagon series on the Left (both green). The light shaded regions represent noise as ${\pm\sigma}$ (one standard deviation) of the sample. Note that the wake region has the greatest standard deviation.}
\end{figure}

The wind tunnel experiments yielded calibration data for each MAST configuration across eight velocities up to 5~m/s and orientation range of $0^\circ\le\theta\le358^\circ$ in $2^\circ$ increments. For a given speed and orientation, each MEMS Hotwire has a unique voltage response. For example, in Figure \ref{fig:subplot_left}, as the MAST rotates counterclockwise from $\theta=0^\circ$, (the orientation at which the A2 sensor is perpendicular to the flow), the MEMS Hotwire rotates out of the flow and thus the voltage climbs due to reduced convection. In the range of approximately $90^\circ<\theta<270^\circ$, the sensor is in the wake of the MAST and has a signal that is independent of angle. As $\theta>270^\circ$, the sensor is rotating back into the flow, and therefore, the voltage decreases to reflect the enhanced convection. Finally, the sensor's voltage returns to its original value once it is again perpendicular to the flow.

A key observation from Figure \ref{fig:subplot_left} is that a single pole-PCB's angular sensitivity is independent of the greater MAST geometry.  A minor exception is the region of $60^\circ<\theta<100^\circ$ which warrants future investigation. The shape of the MAST instead determines the phase offset between adjacent sensors. For example, a hexagon MAST (consisting of six MEMS Hotwires) produces six curves each offset by 60$^\circ$. This is shown in Figure \ref{fig:subplot_right}.

\begin{figure}[h]
\begin{subfigure}{.45\textwidth}
  \centering
    \includegraphics[width=2.1\linewidth]{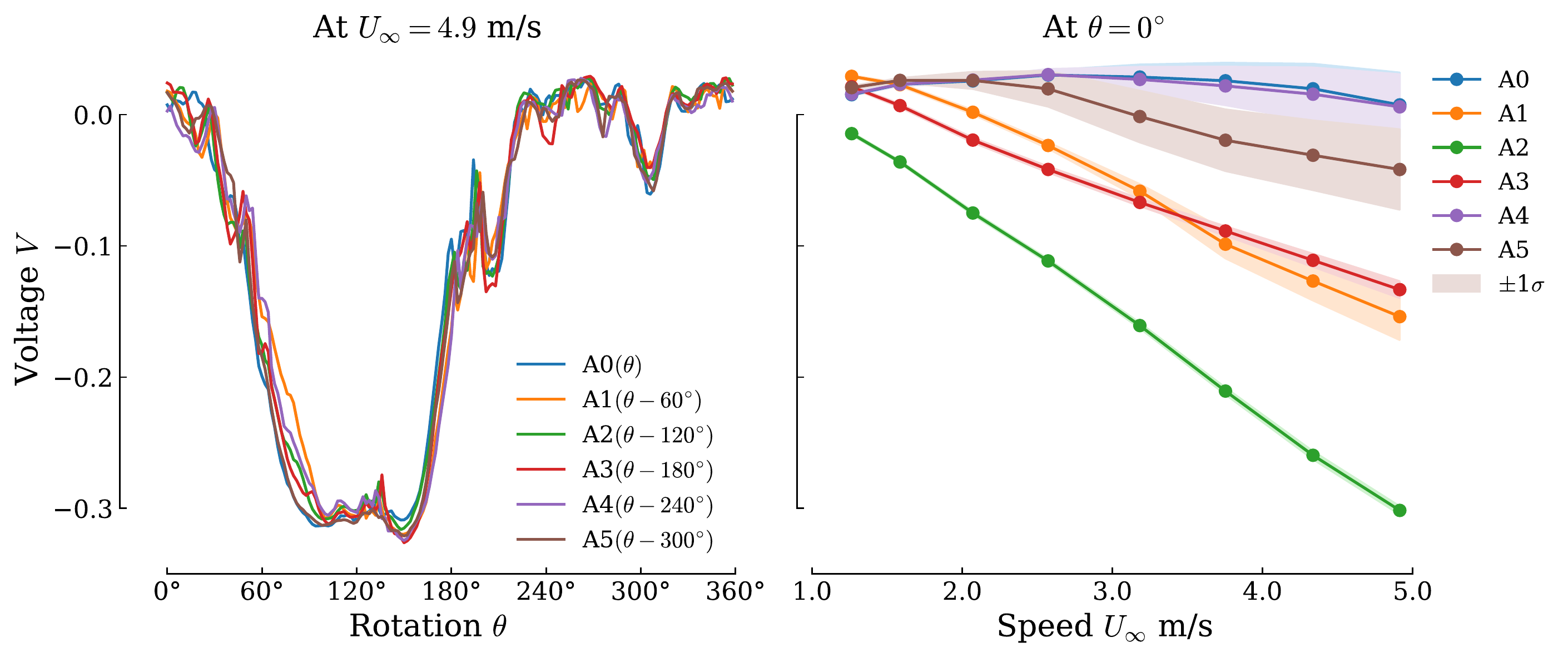}
  \caption{}
  \label{fig:subplot_speed_left}
\end{subfigure}%
\begin{subfigure}[][-168pt][t]{.45\textwidth}
  \centering
  \caption{}
  \label{fig:subplot_speed_right}
\end{subfigure}
\caption{(a) Voltage V vs. MAST rotation $\theta$ at $U_\infty=4.9$~m/s. Analog outputs A1-A5 are shifted to align with A0, which is perpendicular to the flow at $\theta=120^\circ$. This demonstrates the repeatability of the sensor responses. (b) Voltage Vs. Speed at $\theta=0^\circ$. Note that A2 (facing into flow) has the lowest signal. Pairs A1,A3 and A0,A4 are symmetric as expected. A5 faces the wake.}
\end{figure}

The agreement of the six MEMS Hotwire responses can be observed by subtracting the geometric offset of each pole-PCB to align with A0, as shown on Figure \ref{fig:subplot_speed_left}. This collapse occurs at all speeds and illustrates the consistency of the MEMS Hotwire. In addition, the measurements are highly repeatable: in a hysteresis sweep (from $0^{\circ}{\rightarrow}180^{\circ}{\rightarrow}0^{\circ}$ in $0.5^\circ$ increments and at multiple speeds), no hysteresis was detected. Figure \ref{fig:subplot_speed_right} shows the hexagon MAST responses as a function of wind tunnel speed at a particular angle ($\theta=0^\circ$). It is evident that that the MEMS Hotwire perpendicular to the flow at $\theta=0^\circ$ (A2) is both the most sensitive and has the highest signal to noise ratio. The sensor's response when perpendicular to the flow is similar to the performance of a conventional hot-wire, though it does not follow a power law \cite{bruun1995hot}.

Regarding the self-imposed design constraint discussed in Section \ref{sec:MAST}, the fewest number of sensors a MAST geometry needs to resolve a planar flow must be determined. While this is accomplished in the MAST performance comparison of Section \ref{sec:wind-model}, a heuristic can be constructed from the plots to predict the number of sensors required. Each MEMS Hotwire is more sensitive to orientation at certain angles; in these `sensitive regions' noise is minimized and slope is maximized. This region occurs when the sensor is between the perpendicular and wake orientations. In Figure \ref{fig:subplot_left} this is roughly $40-60^\circ$ and $270-320^\circ$, forming a $70^\circ$ region. Dividing the total domain ($360^\circ$) by the sensitive region ($70^\circ$) yields $5.14$, which implies that five or six sensors is the fewest number of sensors whose sensitive regions cover the entire $360^\circ$ domain. This calculation was supported by Figure \ref{fig:subplot_right} where it is noted that one of the six signals is always in the sensitive region (the second lowest signal at any angle).

\begin{figure}
\begin{subfigure}{.45\textwidth}
  \centering
    \includegraphics[width=2.1\linewidth]{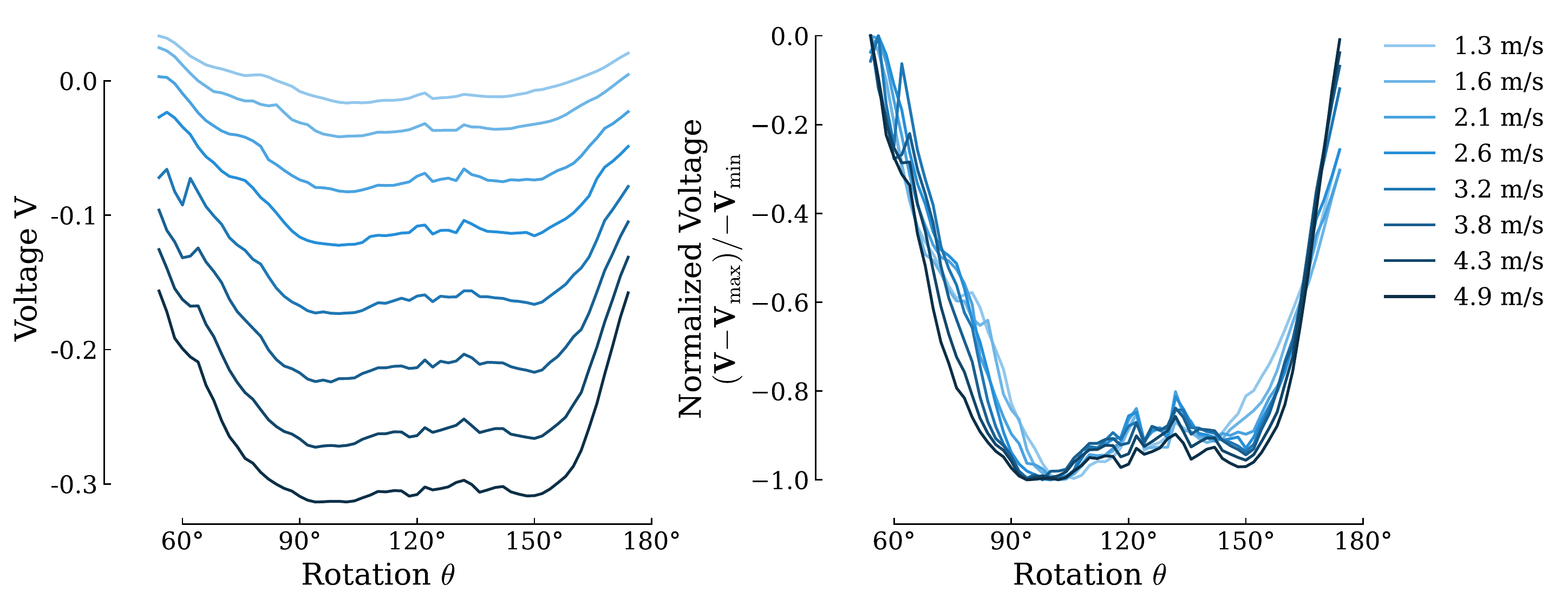}
  \caption{}
  \label{fig:subplot_perp_left}
\end{subfigure}%
\begin{subfigure}[][-155pt][t]{.45\textwidth}
  \centering
  \caption{}
  \label{fig:subplot_perp_right}
\end{subfigure}
\caption{Voltage vs. MAST rotation $\theta$ across the $U_{\infty}$ range tested. (a) The angular response for the A0 pole-PCB is shown. (b) To better show that these responses are differently shaped (angular sensitivity is narrower for lower speeds), the voltage is scaled such that the minimum and maximum of each series is -1.0 and 0.0 respectively.}
\label{fig:subplot_perp}
\end{figure}

To determine flow direction at a static, known speed it would be straightforward to fit a function (e.g., polynomial) to the sensitive regions described above. Onboard the UAV, however, flow speed and direction are both unknown and changing. Furthermore, Figure \ref{fig:subplot_perp} illustrates how angular sensitivity varies with speed; the curves at lower speeds are shaped differently than those at high speeds. Therefore, multiple MEMS Hotwire signals must be input simultaneously into a model that treats both flow speed and direction as unknowns. This model (the functional relationship between the raw signals and the estimated wind vector) is described in Section \ref{sec:wind-model}.

\subsection{Frequency response} \label{sec:frequency-response}

\begin{figure}
\begin{subfigure}{.45\textwidth}
  \centering
    \includegraphics[width=2.1\linewidth]{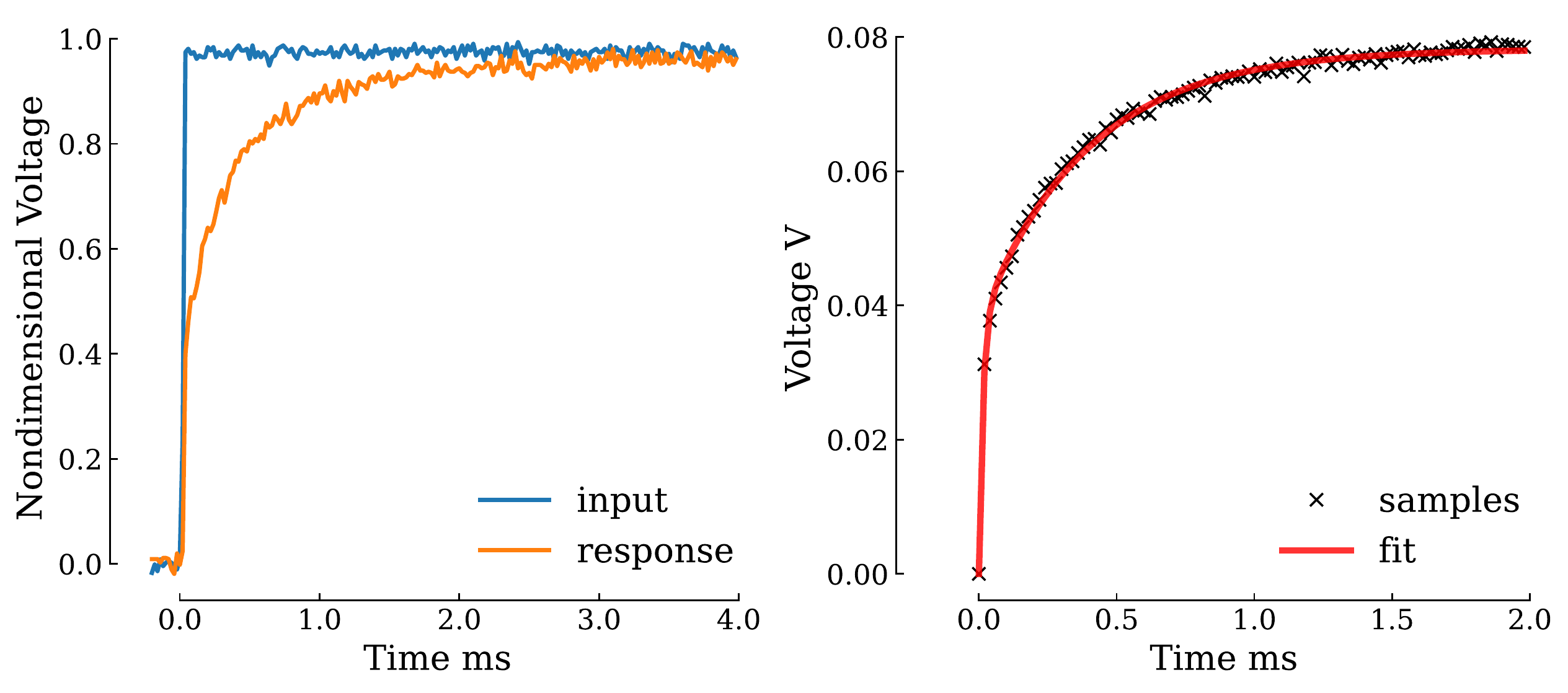}
  \caption{}
  \label{fig:subplot_step_left}
\end{subfigure}%
\begin{subfigure}[][-179pt][t]{.45\textwidth}
  \centering
  \caption{}
  \label{fig:subplot_step_right}
\end{subfigure}
\caption{(a) A single trial of square wave input and sensor response, nondimensionalized between 0 and 1 for comparison. (b) Phase-averaged sensor response ($\mathbf{\times}$-marks) to a 0.1V square wave input over 300 trials. The solid line is the best-fit transfer function (Equation \ref{eqn:ilaplace}).}
\label{fig:subplot_step}
\end{figure}

Finally, the temporal response of the sensor was determined by fitting a transfer function to the phase-averaged step response, which was collected in Section \ref{sec:squareWaveInstrumentation}.
The MEMS Hotwire response is shown in Figure \ref{fig:subplot_step}. In Figure \ref{fig:subplot_step_left}, the nondimensionalized step input and sensor response are shown. There was no measurable lag between the input and the start of the sensor's response. An overdamped transfer function of two real poles and one zero was fit to the first 100 points, or 2~ms.
These points and the $94\%$ fit are shown in Figure \ref{fig:subplot_step_right}. The fit transfer function is:
\begin{equation}
    G(s) = \frac{2.724{\times}10^4s+1.412{\times}10^8}{s^2+7.276{\times}10^4s+1.80{\times}10^8}.
    \label{eqn:transfer}
\end{equation}
This can be rewritten (through inverse Laplace transform) as two first-order systems in series. In particular:
\begin{equation}
V(t) = 0.782271 - 0.373107\exp{(-\omega_1 t)} - 0.409164\exp{(-\omega_2t)},
\label{eqn:ilaplace}
\end{equation}
where the two poles are $\omega_1 = 70.2$ krad/sec and $\omega_2 = 2.57$ krad/sec.

\begin{figure}
    \centering
    \includegraphics[width=0.4\textwidth]{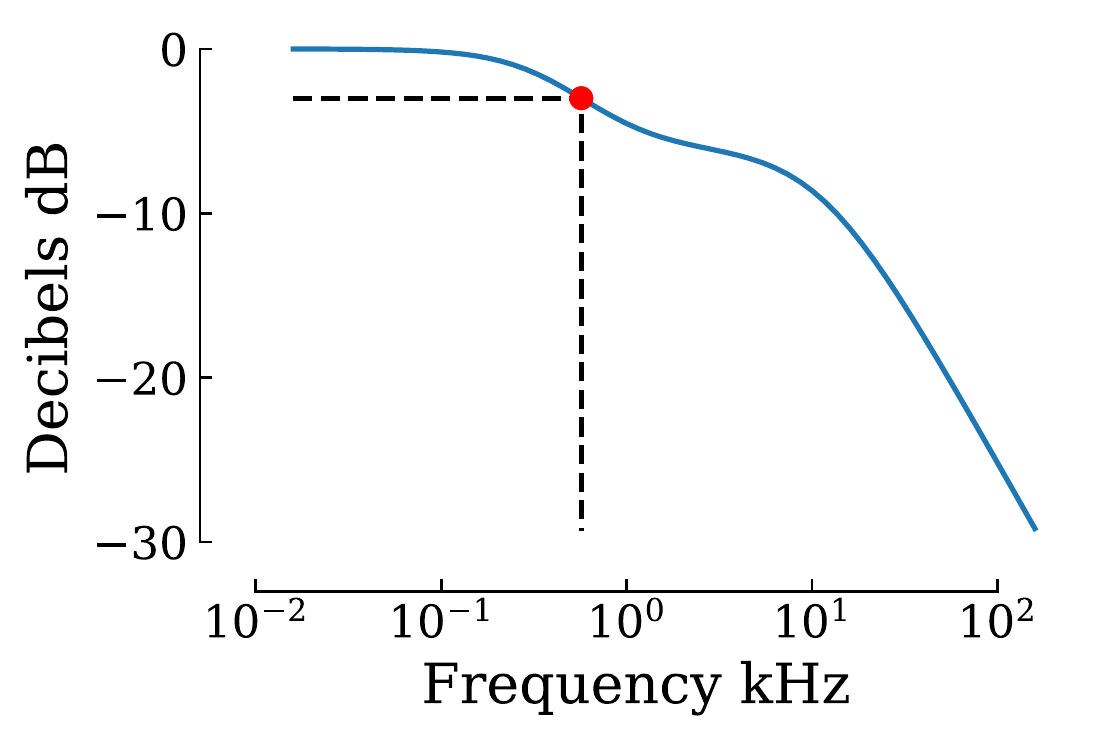}
  \caption{Bode plot. The solid line is the frequency response of the sensor, as determined from the transfer function divided by the DC gain. The dashed line and solid point show the $-3$~dB bandwidth point, which occurs at 570~Hz.}
  \label{fig:bode_plot}
\end{figure}

The rise time of the transfer function is 0.640~ms and its $-3$~dB bandwidth is 570~Hz, as shown by the Bode plot in Figure \ref{fig:bode_plot}. A sensing bandwidth of 570~Hz is sufficiently high for the gust mitigation control problem, as well as for most UAV flow sensing campaigns.

\section{Sensor Model: Methodology and MAST Performance} \label{sec:wind-model}

To effectively use the sensor to improve the drone controller, it is necessary to determine a functional relationship between the raw analog outputs of the sensor and the direction and magnitude of the true wind vector; these real-time estimates of the wind vector could then be logged or used by the control architecture depending on the application. As this nonlinear and complex mapping will be stationary across time, a reasonable approach is to learn the sensor-to-wind functional relationship by training a neural network using the wind tunnel data gathered in Section \ref{subsec:speed-angle-calibration}. This method allows for a quantitative performance comparison of the MAST geometries considered in Section \ref{sec:MAST}, as well as an analysis of the effectiveness of various filtering techniques. Based on the mission requirements for the UAV flow sensing problem, the desiderata of the sensor model are: accurate angle estimation (e.g., $< 5^\circ$ average error), accurate velocity estimation (e.g., $< 0.2$~m/s average error), and compatibility with high-frequency operation.

\begin{table}[h!]
\begin{center}
\begin{tabular}{ c || c c c } 
\diagbox{MAST}{Filter}& \hspace{8mm}$N=1$\hspace{8mm} & \hspace{8mm}$N=2$\hspace{8mm} & \hspace{8mm}$N=5$\hspace{8mm} \\
\hhline{=|===}
 \textbf{Triangle} ($^\circ$) & $13.9$ $(51.4)$& $13.1$ $(47.8)$& $12.1$ $(44.1)$\\
 \hspace{22mm} (m/s) & $0.26$ $(0.85)$ & $0.27$ $(0.87)$ & $0.27$ $(0.79)$ \\ \hline
 \textbf{Square} \hspace{2mm} ($^\circ$) &  $5.3$ $(18.2)$& $5.8$ $(20.1)$& $4.8$ $(16.5)$ \\ 
 \hspace{22mm} (m/s) & $0.16$ $(0.41)$ & $0.16$ $(0.41)$ & $0.15$ $(0.40)$ \\ \hline
 \textbf{Pentagon} ($^\circ$) & $1.6$ $(5.0)$&  $1.8$ $(5.2)$& $1.5$ $(5.2)$\\
 \hspace{25mm} (m/s) & $0.14$ $(0.36)$ & $0.15$ $(0.38)$ & $0.19$ $(0.58)$ \\ \hline
 \textbf{Hexagon} ($^\circ$) & $2.0$ $(6.1)$& $2.1$ $(6.1)$& $1.6$ $(5.0)$\\
\hspace{24mm} (m/s) & $0.11$ $(0.30)$ & $0.12$ $(0.30)$ & $0.11$ $(0.30)$\\
\end{tabular}
\end{center}
\caption{Test error of each MAST-Filter pair for angle [top row, $^\circ$] and magnitude [bottom row, m/s] prediction, with empirical $95\%$-confidence upper bounds on the error in parentheses. Extending the filter length (left to right) weakly improves performance in wind prediction by reducing noise in the data. Increasing the number of sensors is a more beneficial change for estimating wind angle and speed (seen top to bottom).}
\label{table:WindModel}
\end{table}

For each of the twelve MAST-Filter pairs in Table \ref{table:WindModel}, separate wind angle and wind magnitude predictors were trained. The networks were separated for interpretability considerations and to account for the differing complexity of the angle and magnitude prediction problems. The filtering was a pure averaging of $N$ consecutive readings. As such, for any $N$, the ``effective'' sensor model operating frequency is $1000/N$ Hz; $N=1$ corresponds to no filtering, while $N=5$ averages five consecutive readings and only allows the sensor model to operate at 200 Hz. More sophisticated filtering techniques like moving averages (or general finite impulse response (FIR) filters) \cite{DSP_1, DSP_2} were not considered here as they introduce correlations in the data, which violates the assumption of data independence for the learning problem. However, such techniques could improve the sensor performance, and are left for future work.

\subsection{Wind Angle Prediction}
\label{subsec:angle_predict}
For the wind angle estimation problem, the network layer widths were linear in the number of active sensors (see \ref{App:TrainingModels} for details); this scaling was intended to allow each network approximately equal expressivity in order to ensure a fair comparison. The mean angle prediction error for all geometry-filter pairs is summarized in Table \ref{table:WindModel}. The key result is that hexagon and pentagon MAST geometries both greatly outperform the triangle and square geometries as predicted in Section \ref{subsec:speed-angle-calibration}. Furthermore, with pentagon slightly outperforming hexagon, both achieve $\approx 2^\circ$ test error (satisfying the accuracy criterion) in the unfiltered case (satisfying the high-frequency criterion). The empirical angle prediction error distributions for the triangle, pentagon and hexagon geometries are shown on the left in Figure \ref{fig:Hists}.

\subsection{Wind Magnitude Prediction}
\label{subsec:speed_predict}
For the wind magnitude (speed) prediction problem, additional care was required because the sensor was only evaluated at eight distinct wind speeds. While the angle prediction data is ``dense" throughout the entire domain, the wind magnitude data is ``sparse". As such, incautious analysis will underestimate the true prediction error of the learned model by overfitting to the eight particular wind speeds seen in training. 

\begin{figure}[ht]
    \centering
    \includegraphics[width=0.8\textwidth]{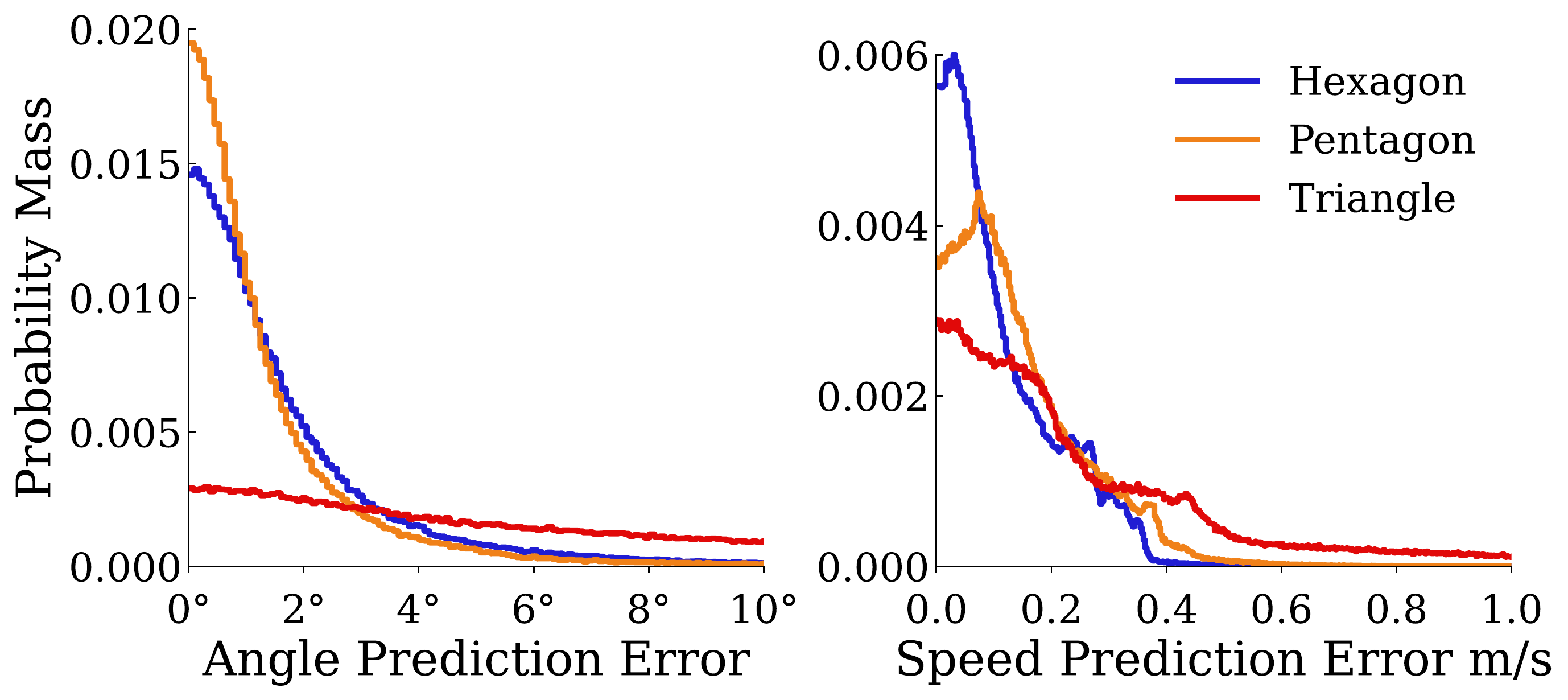}
    \caption{Angle and speed prediction error distributions for hexagon (blue), pentagon (orange), and triangle (red) MAST geometries. Both pentagon and hexagon geometries perform sufficiently well on both tasks.}
    \label{fig:Hists}
\end{figure}

To combat this, two techniques were implemented in parallel. First, the model was simplified by taking as input only the three largest sensor readings for each time step; the inductive bias here is that at most half of the sensors are facing the wind at any given time. For the triangle MAST this procedure utilizes all of the readings, while for the hexagon MAST only half of the readings are used. This allows for a reduction in model size from the angle prediction problem, which discourages overfitting to the sparser data. The second technique utilizes leave-one-out cross-validation (LOOCV) \cite{Murphy_ML} to estimate prediction error on wind speeds unseen during training. This technique incentivizes learning ``less complex" models to promote generalization and reduce overfitting. 

The mean velocity prediction error for all geometry-filter pairs is also summarized in Table \ref{table:WindModel}, and the empirical velocity prediction error distributions for the pentagon and hexagon geometries are shown on the right in Figure \ref{fig:Hists}, with the triangle geometry included for contrast. Again, both pentagon and hexagon geometries attain sufficient overall performance to meet the accuracy criterion for the sensor model, while neither square nor triangle configurations are able to satisfy either of the first two criteria. This demonstrated performance in speed and angle estimation supports the use of five or more sensors to estimate a wind vector.

\begin{figure}[ht]
    \centering
    \includegraphics[width=0.8\textwidth]{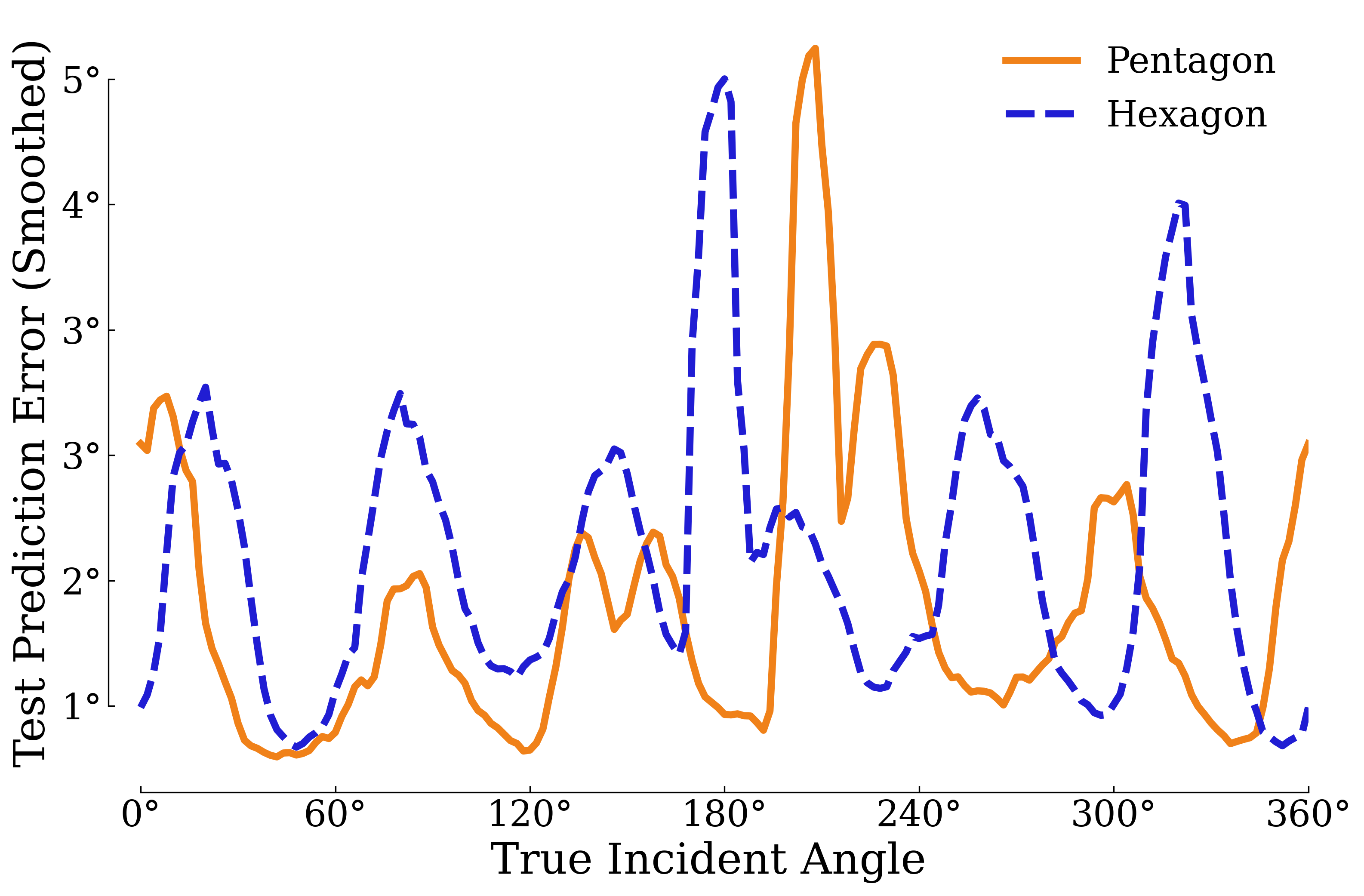}
    \caption{Smoothed mean angle prediction error vs. true incident angle. Note that the pentagon MAST has a peak across zero degrees incidence, giving it only five distinct peaks, whereas hexagon has six peaks, suggesting that observability is geometry-dependent. The reason for the split peak near 180$^\circ$ is unclear; however, because it is observed for both geometries, we hypothesize that it is neural network misclassification due to symmetry. Future work is necessary to determine the underlying mechanism.}
    \label{fig:AngleVsAngleSmooth}
\end{figure}

\subsection{Comparing Pentagon and Hexagon Geometries}
\label{wind:PentvsHex}
To investigate the dependence of angle prediction error upon the true incident wind angle, Figure \ref{fig:AngleVsAngleSmooth} shows the smoothed mean errors in angle prediction for pentagon and hexagon configurations as a function of true incident angle. The smoothing function is a simple moving average with a centered window of length seven. The equivalent plot with the raw prediction error data is included in \ref{App:AngleSmoothing}. Note that both hexagon and pentagon MAST configurations have ``low-observability'' regions where mean errors are larger; interestingly, the number of peaks is equal to the number of sensors, suggesting a cyclic reduction in observability intrinsic to the underlying MAST geometry.

Given the similar performance of the pentagon and hexagon configurations -- both deemed sufficiently accurate -- the pentagon configuration is deemed the `minimal' configuration, using the fewest sensors to achieve the requisite performance. It is notable, and perhaps unintuitive, that the pentagon MAST should outperform the hexagon MAST in angle prediction. However, the sensor models derived here are not shown to be optimal, but merely sufficient to meet desired criteria. With this particular framework, a designer may choose to use the hexagon or pentagon configurations depending on whether greater accuracy in wind speed or wind direction is desired. It is expected that the optimal hexagon sensor model should have the capacity to outperform the optimal pentagon sensor model, but fuller exploration of this is left as future work.
\section{Details for Integration onto a UAV}
\label{sec:uav_integration}

\begin{figure}
  \centering
    \begin{tikzpicture}
        \node[above right, inner sep=0] (image) at (0,0) {
            \includegraphics[width=0.7\textwidth]{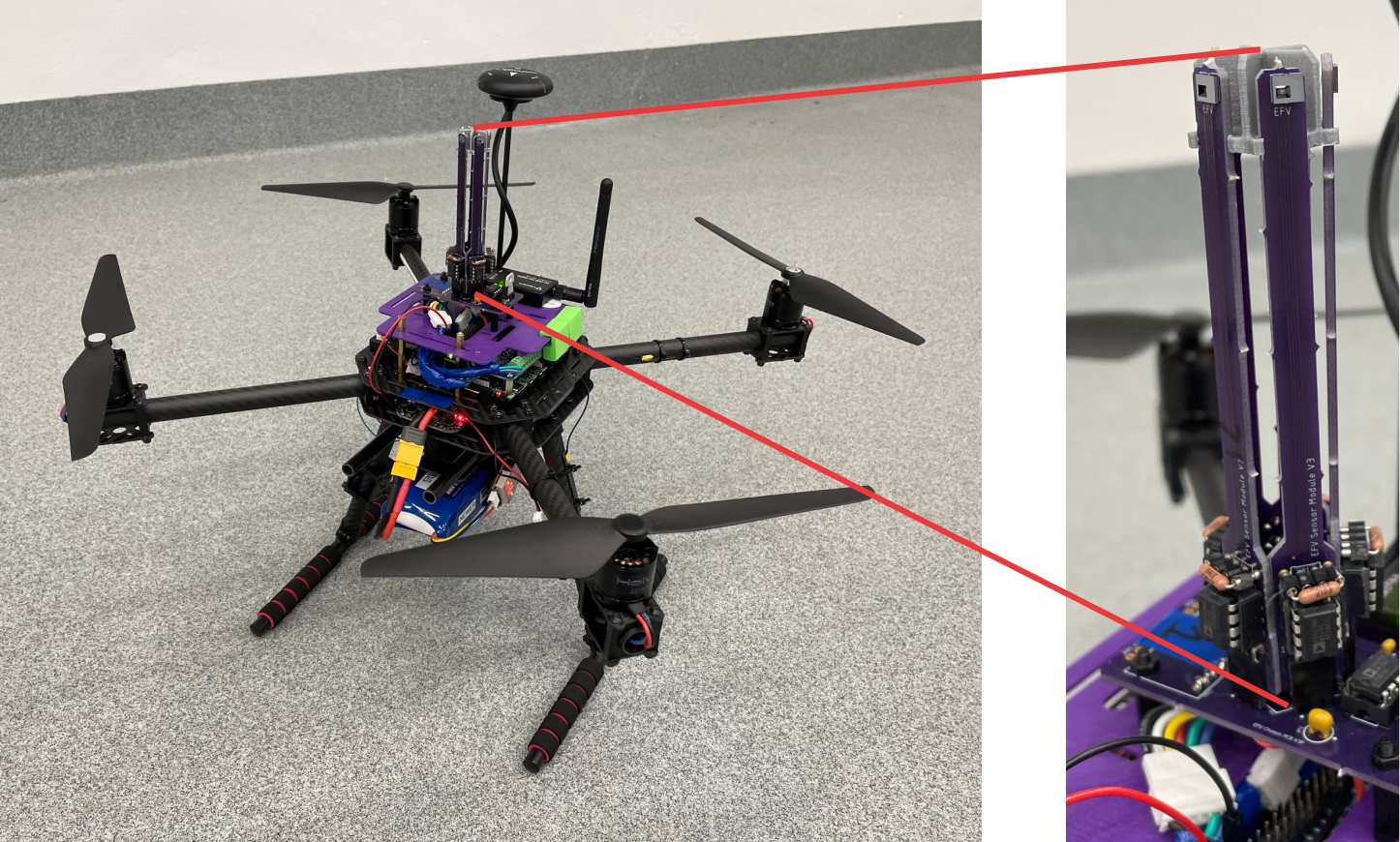}
        };
        \begin{scope}[
        x={($0.1*(image.south east)$)},
        y={($0.1*(image.north west)$)}]
         
            \node[above] at (3.5,10) {FlowDrone};
            \node[above] at (9,10) {MAST};
        \end{scope}
    \end{tikzpicture}
\caption{The FlowDrone (left) is a quadrotor UAV and experimental testbed for MAST integration \cite{flowdrone}. The integrated MAST is shown in greater detail on the right.}
\label{fig:flowdrone_integration}
\end{figure}

A motivation for this work was to develop a sensor that would be suitable for integration onto a drone. As such, the authors have integrated the pentagon MAST onto a Holybro X500 drone (called "FlowDrone"), as shown in Figure \ref{fig:flowdrone_integration}. The authors refer the reader to \cite{flowdrone} for details of the integration and the MAST's performance in-flight. Relevant metrics for the suitability of integration include the MAST's weight, power consumption, and effective sampling frequency. A 16~V power supply provides a constant voltage to the AD587. From this power supply, the MAST draws 0.05~A, indicating that the entire MAST circuit draws 0.8~W. For reference, each of FlowDrone's motors (2216 KV880) draws between 40~W and 100~W at 50\% (hover) and 70\% throttle, respectively. Thus, the MAST has a negligible effect on the drone's power consumption. The heaviest MAST (the hexagon) has a mass of 44~g; the FlowDrone's mass without the hexagon MAST is 1756~g. Therefore, the MAST consists of an increase of only 2.5\% in the drone's mass, rendering the MAST's weight as negligible. It is also noted that, due to semiconductor fabrication techniques, the MEMS Hotwire is very economical, particularly at large volumes. More than one thousand sensors can be made on a single silicon wafer.

Another important criterion for real-time operation is that the sensor model can be executed efficiently enough as not to impede the UAV controller frequency, which can be on the order of 250~Hz. On a Raspberry Pi, the sensor model can be evaluated in 1.56~ms. This corresponds to a maximum wind estimation frequency of 641~Hz, which exceeds the MAST's physical bandwidth of 570~Hz. In other words, the sensor model will affect neither the sensor nor the UAV controller's operational speed. To summarize, the MAST system is well-suited for integration on a UAV and has a negligible negative impact on the drone's capabilities.

\section{Conclusions and Future Work}


The MEMS Hotwire sensor introduced herein provides many of the advantages of hot-wire anemometry --- namely, high spatial and temporal resolution --- in a small and robust package. With a measured bandwidth of 570~Hz, this sensor stands to significantly increase the wind estimation capabilities of UAV platforms through its temporal resolution alone. The pentagon MAST system proposed in this paper strategically places five MEMS Hotwire sensors in the flow to resolve a two-dimensional wind vector (direction and magnitude). While both the pentagon and hexagon configurations both perform sufficiently well (the pentagon outperforms slightly in angle prediction, while the hexagon outperforms slightly in speed prediction), the pentagon arrangement is highlighted as it requires fewer MEMS Hotwire sensors. A neural network sensor model to derive a wind estimate from sensor values is proposed. With its design considerations and demonstrated performance, the MAST system is a unique and promising platform for wind estimation on UAVs, which could be used to improve controller performance or to collect data during field campaigns.

There are limitations of this study that merit future work. One such limitation is that the sensor model assumes that wind is two-dimensional (in-plane). While this is reasonable for the uniform wind-tunnel flow, deployment of the MAST in real-world conditions (e.g., onto a UAV) must consider flow in three-dimensions. A direction for future work is understanding how vertical components of flow affect MAST measurements, and how these components themselves could be measured. In addition, the effects of changing environment temperature were neglected in the wind-tunnel experiments. In reality, a change in environment temperature is expected to induce a change in the steady-state MEMS Hotwire silicon temperature, thus inducing a signal shift at a time scale much slower than the velocity sensing mode. In future studies, a temperature calibration and onboard thermistor can be used to compensate for the effects of changing air temperature and evaluate sensor performance in such environments. Further, there are several aspects of the sensor model that can likely be improved through additional optimization. This includes improving the data filtering and network architecture as well as more general hyperparameter tuning. It is hypothesized that such optimization would validate the intuition that the hexagon MAST outperforms the pentagon MAST in both angle and speed prediction \emph{when the models are near-optimal}. Finally, another line of future work involves integrating the MAST system onto a UAV platform for real-time wind estimation. Such an integration would require an assessment of the effects of the rotors on the MAST's measurements, in particular due to their wake and turbulence. Such a system could then be used in the control loop to improve UAV performance in extreme wind and gusts.


\ack


This work was partially funded by the Air Force Office of Scientific Research grant FA9550-22-1-0020, as well as Princeton's Project X Innovation Fund. N Simon and D Snyder were supported by the National Science Foundation Graduate Research Fellowship Program under Grant No. DGE-2039656. Any opinions, findings, and conclusions or recommendations expressed in this material are those of the author(s) and do not necessarily reflect the views of the National Science Foundation, Air Force Office of Scientific Research, or Princeton University. M Hultmark is co-founder and CEO of Tendo Technologies, Inc., who provided the MEMS Hotwire dies.



\section*{References}
\bibliographystyle{iopart-num-long}
\bibliography{mst.bib}

\newpage{}
\appendix

\section{Detailed Description of the Sensor Model Training}
\label{App:Training}
This Appendix describes all relevant details of the sensor model training procedure, including the data generation, model architectures, and training details. 
\subsection{Sensor Model Training Data}
\label{App:TrainingData}
The training, test, and validation datasets used for the sensor model were sampled at random from subintervals of raw data, utilizing a 15 ms gap between the beginning of each sample to reduce the presence of correlations due to possible transient flow effects. Concretely, for no filtering, every fifteenth data point was selected, whereas for $N=5$, five data points were sampled and averaged, then ten data points were skipped, then the next set of five points were sampled. Thus, the beginning of each sample maintained a distance of fifteen samples, equivalent to 15 ms. 

\subsection{Prediction Model Network Architectures}
\label{App:TrainingModels}
For the wind angle prediction problem, a densely connected network with two hidden layers and ReLU nonlinearity was used for each MAST-filter pair. The input size for each MAST geometry was equal to the number of sensors. For an input layer of width $n_s$ (number of sensors in the MAST geometry), the first hidden layer was of width $8n_s$, and the second hidden layer was of width $4n_s + 5$ (the addition of five was designed to help the smaller geometries). The output was a scalar (the predicted angle of the incoming wind). For the wind speed prediction problem, a densely connected network with one hidden layer and fixed input size was used (again with ReLU nonlinearity). The layer widths \{input, hidden, output\} were $\{3, 6, 1\}$ for all MAST geometries. These architectures were selected by evaluating their performance on the validation data. 

\subsection{Loss Functions and Training Parameters}
\label{App:TrainingParameters}
The cost function was a modified $L_1$ loss that accounted for the `wrapping' of the angle prediction (so that, for example,  predictions of $2\pi - \epsilon_1$ radians were evaluated as being `close' to a true label of $\epsilon_2$ radians. Without this wrapping, gradients would become very large and misdirected near the $0$ / $2\pi$ boundary. The benefit of an $L_1$ loss was direct interpretability of the loss function as the mean absolute error. Losses were evaluated in units of m/s and radians for numerical reasons - this is particularly beneficial if the angle and speed prediction are unified in one network, to ensure that errors are of the same order of magnitude. Results are converted back to degrees for reported results in the main text. 

Training for each network was completed using Pytorch on an Nvidia Titan RTX GPU. The Adam optimizer with default parameters was used for every model. The angle prediction performance of each MAST-geometry pair was evaluated over five random seeds (for a total of 12 MAST-filter pairs by 5 seeds, or 60 models to be trained). Total training time for all 60 of these models took (for the 20-30 epochs sufficient for convergence) on the order of a few hours (note that some time was lost due to inefficiencies in requiring a custom loss function). The speed prediction performance required the training of 480 models (the 60 above times 8 different leave-one-out cross-validation options for each wind speed). Total time for speed prediction training was also 3-4 hours (because only 5 epochs per model were needed for convergence due to the data sparseness, and the default $L_1$ loss could be used for this network). 

\section{Additional Plots for the Sensor Model Performance}
\subsection{Angle Prediction Error Training Curve}
\label{App:AngleTraining}
\begin{figure}[ht]
    \centering
    \includegraphics[width=0.9\textwidth]{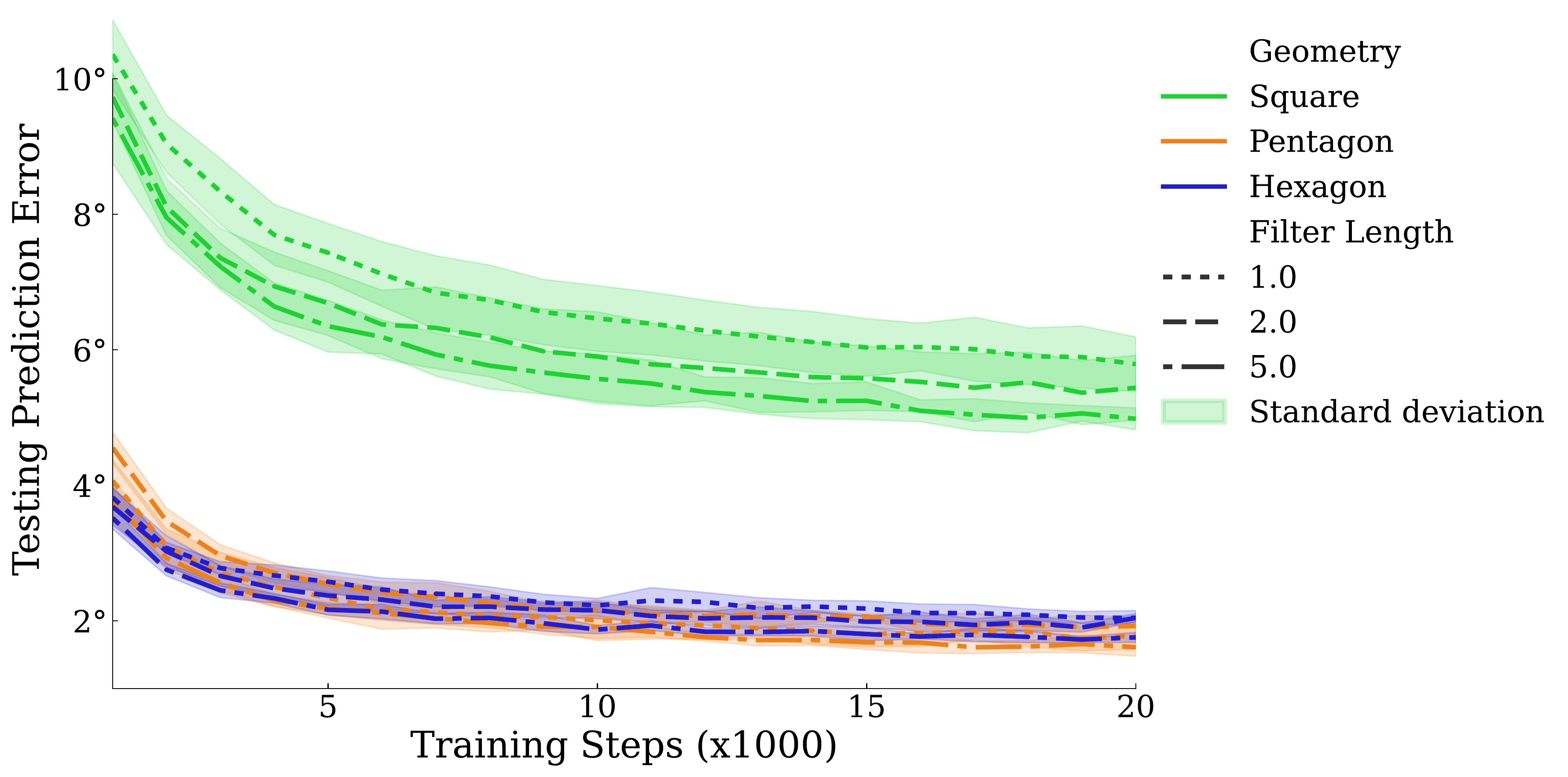}
  \caption{Angular prediction error vs. training iteration for different geometry-filter combinations over the course of training. The hexagon and pentagon MAST geometries both achieve acceptable performance ($< 2^\circ$ test error) in a robust manner (low variance in training curves). The triangle MAST performs quite poorly - see Table \ref{table:WindModel} - and is excluded from this figure for clarity of exposition.}
  \label{fig:trainingcurve}
\end{figure}

Figure \ref{fig:trainingcurve} shows the performance of the three best MAST geometries (Triangle is excluded for scale and clarity; its performance is significantly worse). The confidence bands are relatively tight, suggesting that the procedure was relatively robust.

\subsection{Angle Prediction Error Plot}
\label{App:AngleSmoothing}
This plot shows the true mean angle prediction error against incident angle. Applying the smoothing described in Section \ref{wind:PentvsHex} (centered window of width seven with a simple moving average) yields Figure \ref{fig:AngleVsAngleSmooth}. The smoothed data was shown to highlight more clearly the number of peaks for each geometry. 

\begin{figure}[ht]
    \centering
    \includegraphics[width=0.8\textwidth]{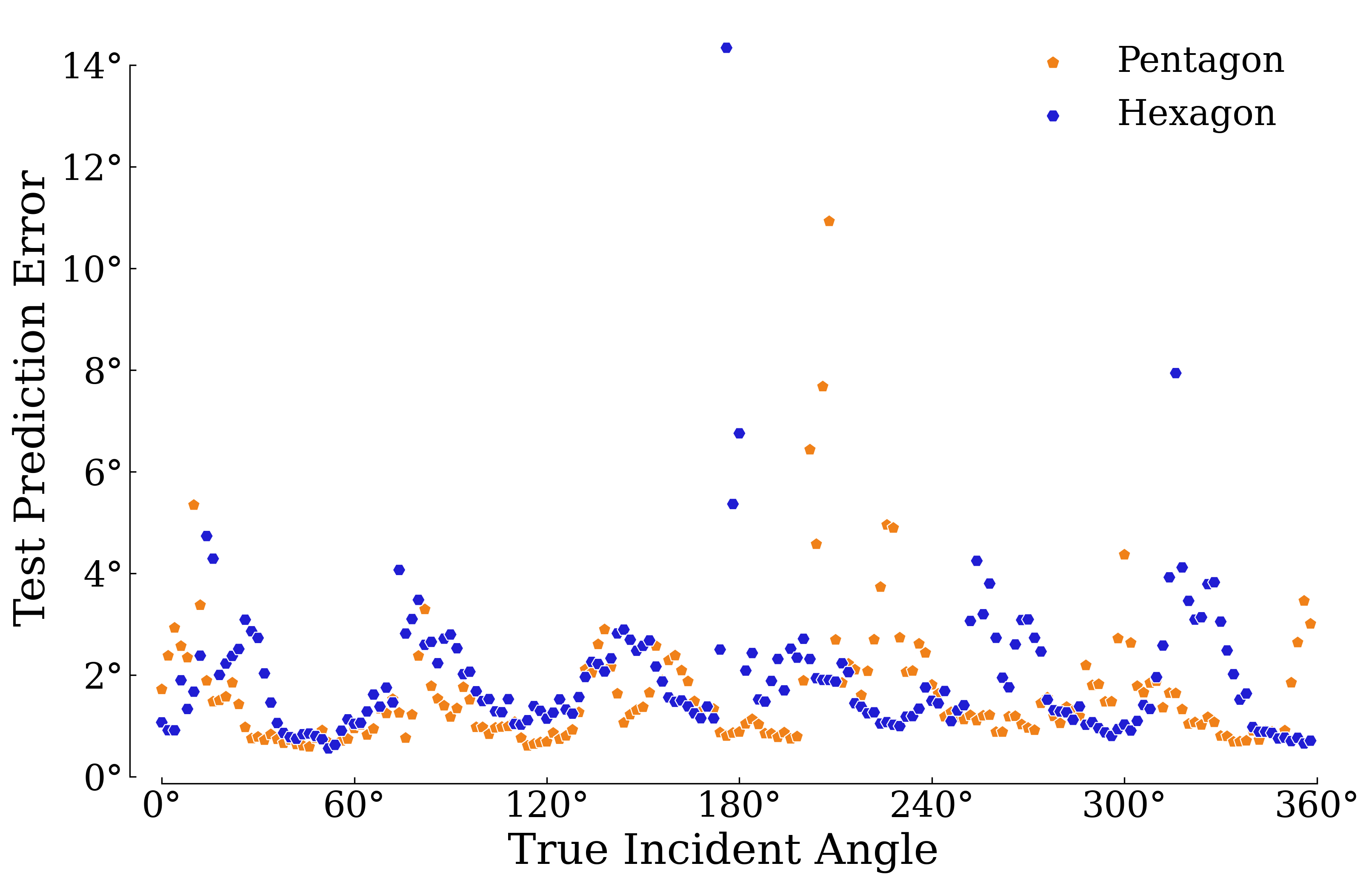}
    \caption{Raw data of mean absolute angle prediction error (deg) vs the true incident angle. The mean of this data across true incident angles yields the performance of each geometry in Table \ref{table:WindModel}. This plot illustrates that incidence ``observability'' varies with the true incident angle in semi-regular ways that depend upon the MAST geometry. }
    \label{fig:AngleVsAngle}
\end{figure}

\end{document}